\documentclass[journal]{IEEEtran}

\usepackage[numbers,sort&compress]{natbib}
\usepackage{url}
\usepackage{bm}
\usepackage{amsfonts}
\usepackage{amsmath}
\usepackage{amssymb}
\usepackage{array,color}
\usepackage{tabularx}
\usepackage{multirow}
\usepackage{booktabs}
\usepackage{amsmath}
\usepackage{makecell}
\usepackage{graphicx}
\usepackage{epstopdf}
\usepackage{graphics}
\usepackage{epsfig}
\usepackage{psfrag}
\usepackage{subfig}
\usepackage{float}
\usepackage{overpic}
\usepackage{stfloats}
\usepackage{float}
\usepackage{amsmath}

\begin{document}
\newcommand{\tabincell}[2]{\begin{tabular}{@{}#1@{}}#2\end{tabular}}

\title{A Correlation Analysis Method for Power Systems Based on Random Matrix Theory}

\author{Xinyi~Xu, Xing~He, Qian~Ai, ~\IEEEmembership{Member,~IEEE,}, Robert~C. Qiu,~\IEEEmembership{Fellow,~IEEE,}
\thanks{This work was supported by the National Natural Science Foundation of China (Grant No. 51577115, 61571296).}}

\maketitle

\begin{abstract}
The operating status of power systems is influenced by growing varieties of factors, resulting from the developing sizes and complexity of power systems; in this situation, the model-based methods need be revisited. A data-driven method, as the novel alternative, on the other hand, is proposed in this paper: it reveals the correlations between the factors and the system status through statistical properties of data. An augmented matrix, as the data source, is the key trick for this method; it is formulated by two parts: 1) status data as the basic part, and 2) factor data as the augmented part. The random matrix theory (RMT) is applied as the mathematical framework. The linear eigenvalue statistics (LESs), such as the mean spectral radius (MSR), are defined to study data correlations through large random matrices. Compared with model-based methods, the proposed method is inspired by a pure statistical approach, without \textit{a prior} knowledge of operation and interaction mechanism models for power systems and factors. In general, this method is direct in analysis, robust against bad data, universal to various factors, and applicable for real-time analysis. A case study, based on the standard IEEE 118-bus system, validates the proposed method.
\end{abstract}


\IEEEpeerreviewmaketitle

\section{Introduction}
\IEEEPARstart{T}{he} operating status of power systems is affected by numerous factors. It is fundamental to understand the statistical correlations between those factors and power systems. These correlations reveal the causes to disturbances and faults \cite{kezunovic2013role}. Nowadays, power systems, large in sizes and complex in structure, are penetrated by more and more various elements, such as distributed generations, flexible loads, and electric vehicles. All these elements lead to strong interaction, multiple coupling, and high randomness in power systems. On this occasion, model-based methods, establishing mechanism models with assumptions and simplifications as essential preconditions, are questionable.

Data have become a strategic resource in power systems. The 4Vs data \cite{ibmfour}, with great potential value, are hard to handle by conventional model-based methods. This situation leads to an emerging paradigm---big data analytics---for power systems. Big data analytics aims to work out statistical features measured by the eigenvalue statistics, without establishing mechanism models. The linear eigenvalue statistics (LESs) are of central interest in statistics \cite{qiu2015smart}. When the matrix size is sufficiently large, LESs tend to deterministic limiting values (expected values). Various forms of the central limit theorems are also established in recent statistical papers. The statistical error between the limiting expected value and the eigenvalue statistics will be decreasing as a function of data size. For more details on the convergence rate, we refer to \cite{o2014central}.

Many studies on big data analytics have been achieved in power systems \cite{xin2014opportunity,zhang20135ws,mayilvaganan2013cloud}. In our previous work, a universal architecture with big data analytics is proposed \cite{he2015big}, and applied successfully in many fields, such as the anomaly detection \cite{he2015unsupervised}, and the 3D power map for situational awareness \cite{he20153d}. This paper is built upon our previous paper \cite{he2015big} in that the augmented matrix contains a remarkable rich statistical information between two (or more) block matrices. The trick is the observation that a matrix of block random submatrices are also a (large) random matrix. This simple observation delivers many interesting results that are useful in large power systems.

\subsection{Contribution}
This paper, based on random matrix theory (RMT), proposes a data-driven method to reveal the correlations between factors and the power system status. An augmented matrix, formed in a certain manner, is presented as the data source. For each factor, the augmented matrix combines status data (as the basic part), on one hand, and factor data (as the augmented part), on the other hand. According to specific researching purposes, status data can be voltages, frequencies, currents and power flows, while factor data can be loads, distributed generation, wind speed, temperature, electricity price, etc. Then, using the big data architecture proposed previously \cite{he2015big}, we conduct real-time analysis based on the augmented matrix, and compare findings with the RMT theoretical predictions (i.e. Ring Law and Marchenko-Pastur Law). During this procedure, the mean spectral radius (MSR), a special case of LESs, is used to indicate data correlations; the kernel density estimation (KDE) is used as an assisted indicator.

In general, the proposed method extracts the correlations in the form of the eigenvalue statistics of measured data. The method involves no knowledge of topologies and parameters of power systems, and is universal to various factors. Besides, the method is robust against random fluctuations in power systems and measuring errors in data. Furthermore, the proposed method is practical for both real-time analysis and off-line analysis, depending on the split-window.

\subsection{Related Work}
Current researches on correlation analysis are mainly model-based methods, for which the mechanism models are essential preconditions. These mechanism models are established based on assumptions and simplifications, and used for specific power systems and factors. Lian studied the effect of dynamic load characteristics on the voltage stability and sensitivity in power systems, using the P--V and the Q--V curves \cite{lian2010voltage}. In Lian's method, the power system is equivalent to an decentralized system; dynamic loads are approximated as differential equations. These processes increase the complexion and inaccuracy of the analysis. Parinya proposed a stochastic stability index to investigate the small signal stability of power systems incorporating wind power \cite{parinya2014study}. The status space equations and energy functions need to be rewritten when the grid changes, and the test system is too small in scale to convince.

Also, some data-driven methods for correlation analysis are proposed recently, such as the principal components analysis, the artificial neural networks, the support vector machine \cite{xie2014dimensionality}. Eltigani utilizes artificial neural networks (ANNs) in assessing the transient stability \cite{eltigani2013implementation}. In his approach, the power system is described by an equivalent single machine infinite bus system, which cannot reflect accurately the actual state of the system. Moreover, with the scale-up of the system and increase of training samples, the training speed of ANNs progressively slows down.

\section{Random Matrix Theory and Data Processing}
The frequently used notations in this section are given in Table.~\ref{Notation-1}.
\begin{table}[ht]
\centering
\caption{Notations for RMT and data processing}
\label{Notation-1}
\begin{tabular*}{8.8cm} {ll}
\toprule[1pt]
\textbf{Notations} & \textbf{Means}\\
\hline
$\mathbf X$, $\mathbf x$, $x$, $x_{i,j}$ & a matrix, a vector, a single value, an entry of a matrix \\
$\mathbf X^H$, $\mathbf x^H$ & transpose of a matrix and a vector \\
$\mu(\mathbf x)$, $\sigma^2(\mathbf x)$ & mean, variance for $\mathbf x$ \\
$\mathbf{\Omega}$ & the data source \\
$\mathbb C^{N \times T}$ & $N \times T$ dimensional complex space \\
$N$, $T$ & the row size and the column size of the split window \\
$\hat{\mathbf X}$ & a raw data matrix \\
$\widetilde{\mathbf X}$ & a standard non-Hermitian matrix \\
$\mathbf X_u$ & the singular value equivalent of $\mathbf X$ \\
$\hat{\mathbf Z}$ & the matrix product \\
$\widetilde{\mathbf Z}$ & the standard matrix product \\
$\lambda_{\mathbf Z}$ & eigenvalues of $\mathbf Z$ \\
$|\lambda|$ & radius of eigenvalue $\lambda$ on the complex plane \\
$\kappa_{\text{MSR}}$ & the mean spectral radius \\
$\mathbf S$ & the sample covariance matrix \\
\hline
\end{tabular*}
\end{table}

\subsection{Random Matrix Theory}
The random matrix theory (RMT), developed from several different sources in the early 20th century, is one of the statistical foundations for big data analytics. It is used as an important mathematical tool in various fields, namely, physics, finance, wireless communication engineering, etc.

Massive data can be naturally represented by large random matrices \cite{qiu2014cognitive}. The random matrix model is the most general: rectangular and complex. In our formulation, we view the $N$ variables as space samples of a random network (or graph). For each variable, we make $T$ observations. As a result, a random matrix of $N \times T$ is obtained as our data matrix, which is the starting part for our analysis.

According to RMT, when the dimensions of a random matrix are sufficiently large, the empirical spectral distribution (ESD) of its eigenvalues converges to some theoretical limits, such as Ring Law and Marchenko-Pastur Law (M-P Law) \cite{zhang2014data}. For details on the Ring Law and M-P Law, please refer to Appendix A. It is noted that although the asymptotic convergence in RMT is considered under infinite dimensions, the asymptotic results are remarkably accurate for relatively moderate matrix sizes such as tens. This is the very reason why RMT can be used in practical world.

\subsection{Real-time Data Processing for RMT}
Currently, there exists no general standardized definition for big data. In this paper, we use a mathematical definition proposed in our previous work \cite{he2015big}. In a power system, assume that there are $n$ kinds of measurable status variables. At the sampling time $t_i$, measured data of these variables are formed as a column vector $\hat{\mathbf x}(t_i) \! = \! (\hat x_1,\hat x_2,\ldots,\hat x_n)^H$ \cite{donoho2000high}.  For a series of time, we can arrange these vectors ${\hat {\bf x}} $ in chronological order to form a matrix as a data source $\mathbf{\Omega}$  for further analysis (${\hat {\bf x}}\in \mathbf{\Omega}$).

Within  $\mathbf{\Omega},$ we can obtain a raw data matrix $\hat{\mathbf X} \in \mathbb C^{N \times T}$ arbitrarily by using a split-window. Then, we convert it into a standard non-Hermitian matrix $\widetilde{\mathbf X}$ with following algorithms.
\begin{equation}
\widetilde x_{i,j} \! = \! (\hat x_{i,j} \! - \! \mu (\hat{\mathbf x}_i)) \! \times \! \frac{\sigma(\widetilde{\mathbf x}_i)}{\sigma(\hat{\mathbf x}_i)} \! + \! \mu (\widetilde{\mathbf x}_i)
\end{equation}
where $\hat{\mathbf x}_i\!=\!(\hat x_{i,1},\hat x_{i,2},\ldots,\hat x_{i,T})$, $\mu (\widetilde{\mathbf x}_i)\!=\!0$, $\sigma(\widetilde{\mathbf x}_i) \! = \! 1$ for $i\!=\!1,2,\!\ldots\!,N$ and $j\!=\!1,2,\!\ldots\!,T$. The matrix $\widetilde{\mathbf X}_u \! \in \! \mathbb C^{N \times N}$ is introduced as the singular value equivalent of $\widetilde{\mathbf X}$ by
\begin{equation}
\widetilde{\mathbf X}_u=\sqrt {\widetilde{\mathbf X} \widetilde{\mathbf X}^H} \mathbf U
\end{equation}
where $\mathbf U \! \in \! \mathbb C^{N \times N}$ is a Haar unitary matrix and $\widetilde{\mathbf X}_u \widetilde{\mathbf X}_u^H \! = \! \widetilde{\mathbf X}\widetilde{\mathbf X}^H$.

For multiple arbitrarily assigned standard non-Hermitian matrices $\widetilde{\mathbf X}_i~(i\!=\!1,2,\ldots,L)$, the matrix product is obtained by $\hat{\mathbf Z} \! = \! \prod_{i=1}^{L} \widetilde{\mathbf X}_{u,i}$. Then, $\hat{\mathbf Z}$ is converted to the standard matrix product $\widetilde{\mathbf Z}$ by
\begin{equation}
\widetilde{\mathbf z}_i=\frac{\hat{\mathbf z}_i}{\sqrt{N}\sigma(\hat{\mathbf z}_i)}
\end{equation}
where $\widetilde{\mathbf z}_i \! = \! (\widetilde z_{i,1},\widetilde z_{i,2},\ldots,\widetilde z_{i,N})$ and $\hat{\mathbf z}_i \! = \! (\hat z_{i,1},\hat z_{i,2},\ldots,\hat z_{i,N})$.

For the standard matrix product $\widetilde{\mathbf Z}$, the sample covariance matrix is obtained by
\begin{equation}
\mathbf S\!=\!\frac{1}{N}\mathbf Y \mathbf Y^H\!=\!\widetilde{\mathbf Z}\widetilde{\mathbf Z}^H
\end{equation}
where $\mathbf Y=\sqrt{N}\widetilde{\mathbf Z}$, and $\sigma^2(\mathbf y_i)=\sigma^2(\sqrt{N}\widetilde{\mathbf z}_i)=1$.

In order to conduct real-time analysis, we use a specific split-window to obtain the raw data matrix $\hat{\mathbf X}$ from $\mathbf{\Omega}$, namely, the real-time split-window. The real-time split-window truncates measured data at continuous sampling times, where the last sampling time is the current time. In other words, at the sampling time $t_i$, the raw data matrix $\hat{\mathbf X}_{t_i}$ is formed by
\begin{equation}
\hat{\mathbf X}(t_i)=(\hat{\mathbf x}(t_{i-T+1}),\hat{\mathbf x}(t_{i-T+2}),\ldots,\hat{\mathbf x}(t_i))
\end{equation}
where $\hat{\mathbf x}(t_j) \! = \! (\hat x_1,\hat x_2,\ldots,\hat x_N)^H$ is measured data at the sampling time $t_j$.

The data processing procedure above is organized as following steps. The standard matrix product $\widetilde{\mathbf Z}$ is calculated for Ring Law; the sample covariance matrix $\mathbf S$ is calculated for M-P Law. For simplicity, we set $L \! = \! 1$ in the matrix product $\hat{\mathbf Z}$.
\begin{table}[h]
\centering
\begin{tabular*}{8.8cm} {l}
\toprule[1pt]
\textbf{Steps of Data Processing for Ring Law}\\
\hline
\tabincell{l}{1) At the sampling time $t_i$, obtain the raw data matrix $\hat{\mathbf X}(t_i) \! \in \! \mathbb C^{N \times T}$} \\
from the data source $\mathbf {\Omega}$. \\
2) Convert $\hat{\mathbf X}(t_i)$ into the standard non-Hermitian matrix $\widetilde{\mathbf X}(t_i)$. \\
3) Calculate the singular value equivalent $\widetilde{\mathbf X}_u(t_i)$ of $\widetilde{\mathbf X}(t_i)$. \\
4) Form the matrix product $\hat{\mathbf Z}(t_i)=\widetilde{\mathbf X}_u(t_i)$. \\
5) Convert $\hat{\mathbf Z}(t_i)$ into the standard matrix product $\widetilde{\mathbf Z}(t_i)$. \\
6) Calculate the sample covariance matrix $\mathbf S(t_i)=\widetilde{\mathbf Z}(t_i)\widetilde{\mathbf Z}(t_i)^H$ \\
\hline
\end{tabular*}
\end{table}

\subsection{Linear Eigenvalue Statistics and Kernel Density Estimation}
\subsubsection{Linear Eigenvalue Statistics}

\

The linear eigenvalue statistics (LESs), major focuses in this paper, indicate the statistical characteristics of large random matrices. The linear eigenvalue statistic of a random matrix $\mathbf X$ is defined as \cite{jana2014fluctuations}
\begin{equation}
\mathcal{N}_n(\varphi)=\sum_{i=1}^n\varphi(\lambda_i)
\end{equation}
where $\lambda_i~(i\!=\!1,2,\!\ldots\!,n)$ are eigenvalues of $\mathbf X$, and $\varphi(\cdot)$ is a test function.

The mean spectral radius (MSR), a special case of LESs, is used to reflect the eigenvalue distribution of the standard matrix product $\widetilde{\mathbf Z}$; it is defined as the mean distribution radius of eigenvalues, formulated by
\begin{equation}
\kappa_{\text{MSR}}=\frac{1}{N}\sum_{i=1}^N\vert \lambda_{\widetilde{\mathbf Z},i}\vert
\end{equation}
where $\lambda_{\widetilde{\mathbf Z},i}~(i\!=\!1,2,\!\ldots\!,N)$ are eigenvalues of $\widetilde{\mathbf Z}$, and $\vert\lambda_{\widetilde{\mathbf Z},i}\vert$ is the radius of the eigenvalue $\lambda_{\widetilde{\mathbf Z},i}$ on the complex plane. Based on MSR, we define the variance of spectral radius (VSR) as a further reflection of eigenvalue distribution; it reflects the dispersion degree of the eigenvalues of $\widetilde{\mathbf Z}$, formulated by
\begin{equation}
\kappa_{\text{VSR}}=\frac{1}{N-1}\sum_{i=1}^N (\vert \lambda_{\widetilde{\mathbf Z},i}\vert-\kappa_{\text{MSR}})^2
\end{equation}
Note that since these complex eigenvalues $\lambda_{\widetilde{\mathbf Z},i}~(i\!=\!1,2,\!\ldots\!,N)$ are highly correlated random variables (each is a complicated matrix function of random matrices $\widetilde{\mathbf X}_i~(i\!=\!1,2,\!\ldots\!,L)$, $\kappa_{\text{MSR}}$ and $\kappa_{\text{VSR}}$ are both random variables.

\subsubsection{Kernel Density Estimation}

\

The kernel density estimation (KDE) depicts the ESD of the sample covariance matrix $\mathbf S$ by following PDF
\begin{equation}
f_{\text{KDE}}(\lambda_{\mathbf S})=\frac{1}{Nh}\sum_{i=1}^N\operatorname{K}(\frac{\lambda_{\mathbf S}-\lambda_{\mathbf S,i}}{h})
\end{equation}
where $\lambda_{\mathbf S,i}~(i\!=\!1,2,\!\ldots\!N)$ are eigenvalues of $\mathbf S$, and $\operatorname{K}(\cdot)$ is the kernel function for the bandwidth parameter $h$.

\section{A Correlation Analysis Method for Power Systems based on Augmented Matrices}
The frequently used notations are given in Table.\ref{Notation-2}.
\begin{table}[h]
\centering
\caption{Notations for correlation analysis}
\label{Notation-2}
\begin{tabular*}{8.8cm} {ll}
\toprule[1pt]
\textbf{Notations} & \textbf{Means}\\
\hline
$n$ & the number of status variables of power systems\\
$m$ & the number of influential factors in power systems\\
$t$ & the study duration \\
$t_s$ & the sampling time \\
$\mathbf B$ & the status matrix \\
$\mathbf c$ & a factor vector \\
$\mathbf D$ & the matrix duplicating $\mathbf c$ for $k$ times \\
$\mathbf E$ & the noise matrix \\
$m_e$ & the magnitude of white noise \\
$\mathbf C$ & the factor matrix \\
$\rho$ & the signal-to noise ratio \\
$\mathbf A$ & the augmented matrix \\
\hline
\end{tabular*}
\end{table}

\subsection{Augmented Matrix Method for Power Systems}
Different factors have different effects on power systems in their own ways. According to big data analytics, the relationships between factors and power systems can be revealed by data correlations. Based on RMT, it is feasible to extract correlations from a data source including both status data and factors data, namely, augmented matrix.

In a power system, assume that there are $n$ types of status variables and $m$ types of influential factors, both of which are measurable. In a study period $t_i~(i\!=\!1,2,\!\ldots\!,t)$, measured data of each factor are formed as a row vector $\mathbf c_j \in \mathbb C^{1 \times t} ~(j\!=\!1,2,\!\ldots\!,m)$ (i.e. factor vector), and measured data of status variables for the power system are formed as a matrix $\mathbf B\!\in\!\mathbb C^{n \times t}$ (i.e. status matrix).

In order to balance the proportion of factors data and status data in the data source, we form a factor matrix for each factor vector. First, for the factor $\mathbf c_j$, we duplicate it for $k$ times to construct a matrix $\mathbf D_j$, formulated by
\begin{eqnarray}
\mathbf D_j=\left[
\begin{array}{c}
\mathbf c_j \\
\mathbf c_j \\
\vdots \\
\mathbf c_j \\
\end{array}
\right]_{k \times t}\quad(j=1,2,\ldots,m)
\end{eqnarray}
where $k$ has the similar size with $n$. Then, we introduce white noise into $\mathbf D_j$ to reduce the correlations among its own rows. The noise matrix is $\mathbf E = \{ e_{i,j} \}_{k \times t}$, where $e_{i,j}$ is random variable according with the normal distribution. Finally, the factor matrix is formulated by
\begin{equation}
\mathbf C_j=\mathbf D_j+m_{e,j} \mathbf E \quad (j=1,2,\ldots,m)
\end{equation}
where $m_{e,j}$ is the magnitude of white noise for the factor matrix $\mathbf C_j$. The signal-to-noise ratio (SNR) of the factor matrix $\mathbf C_j$ is defined as
\begin{equation}
\rho_j=\frac{\operatorname{Tr} (\mathbf D_j \mathbf D_j^H)}{\operatorname{Tr}(\mathbf E \mathbf E^H)\times m_{e,j}^2}\quad(j=1,2,\ldots,m)
\end{equation}
where $\operatorname{Tr}(\cdot)$ is the trace of the matrix. The value of $\rho_j$, requiring careful selections, will affect the performance of correlation analysis. In this paper, we set the same SNR for all factors, to ensure the consistency of correlation analysis. Therefore, for each $\mathbf D_j(j\!=\!1,2,\!\ldots\!,m)$, the value of $m_{e.j}$ is calculated by
\begin{equation}
m_{e,j}=\sqrt{\frac{\operatorname{Tr}(\mathbf D_j\mathbf D_j^H)}{\operatorname{Tr}(\mathbf E\mathbf E^H) \times \rho}}\quad(j=1,2,\ldots,m)
\end{equation}

For each factor $\mathbf c_j$, we can construct an augmented matrix for parallel correlation analysis, formulated by
\begin{eqnarray}
\mathbf A_j=\left[
\begin{array}{c}
\mathbf B \\
\mathbf C_j
\end{array}
\right]\quad (j=1,2,\ldots,m)
\end{eqnarray}

\subsection{Status Data and Factor Data}
\subsubsection{Status Variables of Power System}

\

The operating status of power systems can be estimated by various types of status variables, such as frequencies, voltages, currents and power flows. One or more types of these status variables can form the status matrix $\mathbf B$. In this paper, we use magnitudes of bus voltages as status data for the following considerations:\\
\small{}
a) The voltage magnitude is one of the most basic parameters in power systems. It is measurable and available on every bus with common measuring devices. Therefore, magnitudes of bus voltages have considerable redundancy and accuracy for correlation analysis.\\
b) The voltage magnitude does not involve the topology of power systems. Therefore, we can conduct analysis without  the \textit{a prior} knowledge of network structures and parameters.\\
\normalsize{}
\subsubsection{Factors in Power Systems}

\

The operating status of power systems is mainly affected by electrical factors, climatic factors and economic factors.\\
\small{}
a) Electrical factors--nodal loads and distributed generations, etc.\\
b) Climatic factors---temperature, wind speed, light intensity, etc.\\
c) Economic factors---electricity price, gross domestic product, etc.\\
\normalsize{}

Since data normalization is used during data processing, factor data and status data in the augmented matrix can have different units and magnitudes. In consideration of different sampling frequencies for status data and factor data, it can be assumed that data with lower sampling frequency are the same values in a sampling interval.

\subsection{Correlation Analysis Method and Its Advantages}
Based on RMT and the augmented matrix method, a correlation analysis method for power systems is designed as following steps.
\begin{table}[h]
\centering
\begin{tabular*}{8.8cm}{l}
\toprule[1pt]
\textbf{Steps of Correlation Analysis for Power Systems} \\
\hline
\tabincell{l}{1) In a study period, form the status matrix $\mathbf B \! \in \! \mathbb C^{n \times t}$} and factor vectors  \\
$\mathbf c_j \! \in \! \mathbb C^{1 \times t} ~(j \! = \! 1,2,\ldots,m)$. \\
2) At each sampling time, \\
\quad 2a) acquire the standard matrix product $\widetilde{\mathbf Z}$ from $\mathbf B$. \\
\quad 2b) calculate $\kappa_{\text{MSR}}$ and $\kappa_{\text{VSR}}$ of $\widetilde{\mathbf Z}$.\\
3) Meanwhile, for each factor $\mathbf c_j$, construct the augmented matrix $\mathbf A_j$. \\
4) At each sampling time, \\
\quad 4a) acquire the standard matrix product $\widetilde{\mathbf Z}$ from $\mathbf A_j$. \\
\quad 4b) calculate $\kappa_{\text{MSR}}$ and $\kappa_{\text{VSR}}$ of $\widetilde{\mathbf Z}$. \\
\quad 4c) acquire the sample covariance matrix $\mathbf S$ from $\widetilde{\mathbf Z}$. \\
\quad 4d) calculate $f_{\text{KDE}}$ of $\mathbf S$ and draw the $f_{\text{KDE}}-\lambda$ curve. \\
\tabincell{l}{5) Draw $\kappa_{\text{MSR}}\!-\!t$ curves of the status matrix and the augmented matrix} \\
for each factor. \\
\hline
\end{tabular*}
\end{table}

Step 2) is conducted for the anomaly detection, where the $\kappa_\text{MSR}-t$ curve of the status matrix discovers the signals in power systems \cite{he2015unsupervised}. Step 3)--4), analyzing the correlations between the system status and each factor, aim to determine the causes to the signals. During the analysis procedure, $\kappa_{\text{MSR}}$ and $f_{\text{KDE}}$ are calculated as correlation indicators.

The proposed correlation analysis method is driven by measured data, and based on statistical theories. The procedure involves no mechanism models for power systems and factors; it eliminates the interference brought by assumptions and simplifications. Compared with model-based methods, the proposed method is data-driven, and universal for various factors. Meanwhile, the proposed method has strong robustness against random fluctuations in systems and measuring errors in data. Besides, the method is practical for real-time analysis by using a specific split-window. The advantages above will be verified in case studies.

\section{Case Studies}
The proposed method is tested with simulated data in the standard IEEE 118-bus system. Detailed information of the system is referred to the case118.m in Matpower package and Matpower 4.1 User's Manual \cite{zimmerman2011matpower}. In the simulations, we regard the active load of each bus as a factor; each change of a factor is considered as a signal.

Four cases are designed in different scenarios to validate the effectiveness of the proposed method. In case 1, case 2 and case 3, three kinds of signals are added into single factor to affect the operating status of the test system. In case 4, multiple factors with different kinds of signal produce effects on the system status.

In order to determine the causes to the signals, simulated data of each factor are used to conduct correlation analysis. It is a parallel analysis procedure, and we can only pay our attention to potential factors. Here, we just illustrate the results with the load of bus 117 and bus 54 to show the performance of the proposed method.

Assume that status data are sampled at each time interval, while factor data are sampled every 50 time intervals. Besides, white noise is introduced into both status data and factor data to represent random fluctuations and measuring errors.

For all cases, let $n\!=\!118,t\!=\!1000,N\!=\!118,T\!=\!240,k\!=\!\frac{1}{2}N\!=\!59,\rho\!=\!500$.

\subsection{Correlation Analysis for Single Factor}
\subsubsection{Case 1 \!-\! Step signal in the load of bus 117}

\

\begin{figure*}[ht]
\centering
\subfloat[Data source: status matrix]{
\includegraphics[width=0.32\textwidth]{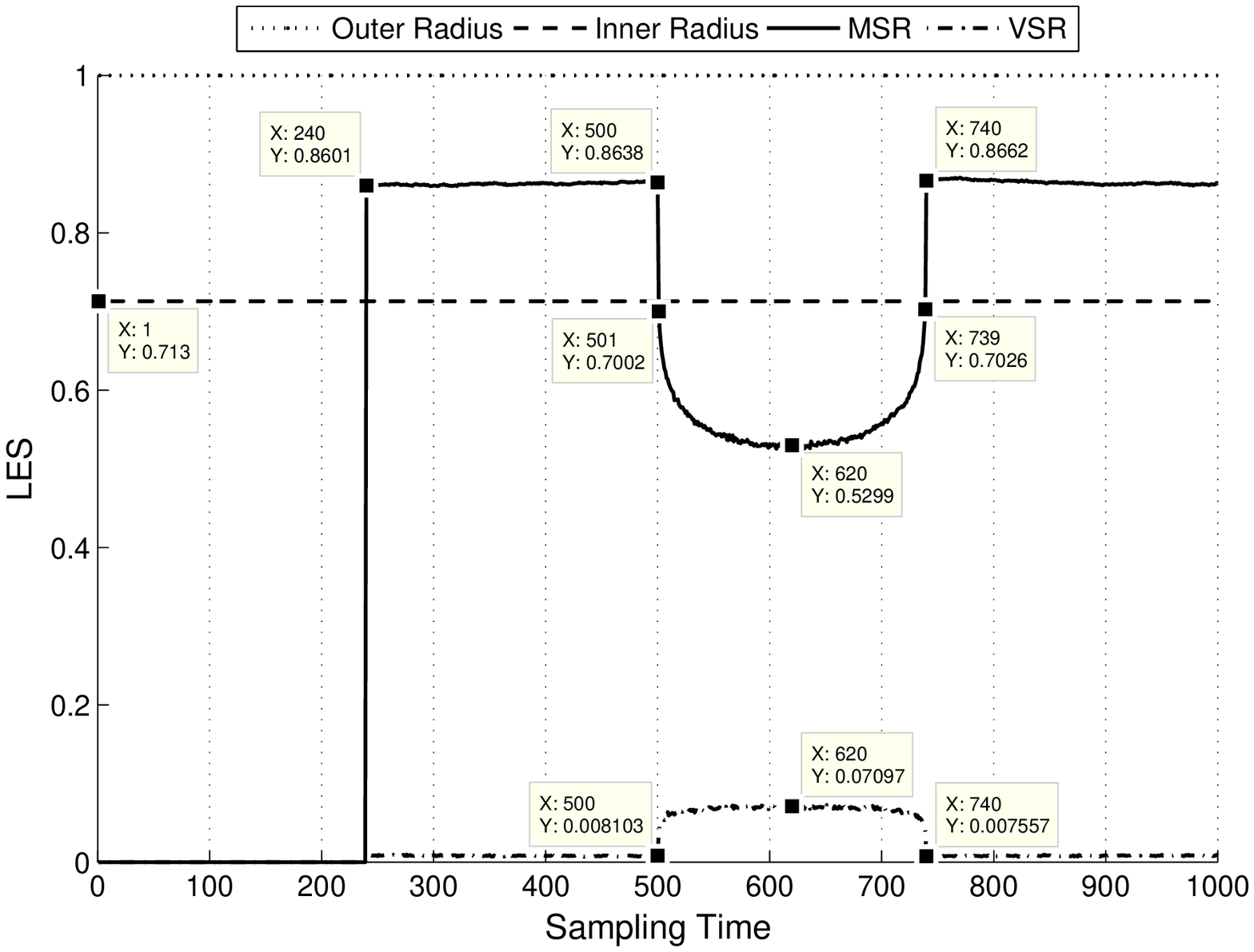}
\label{1-0}
}
\subfloat[Data source: augmented matrix including load data of bus 117]{
\includegraphics[width=0.32\textwidth]{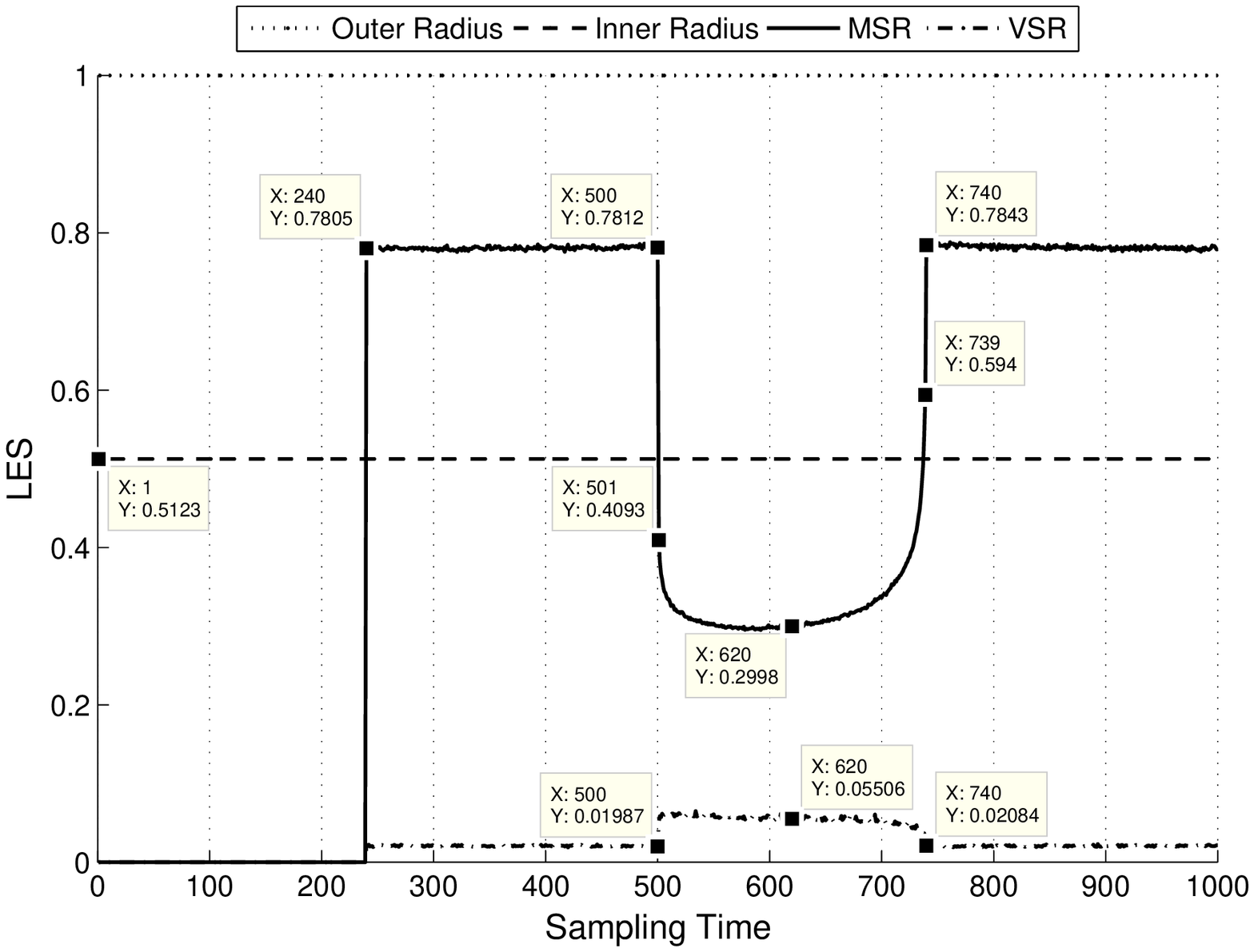}
\label{1-117}
}
\subfloat[Data source: augmented matrix including load data of bus 54]{
\includegraphics[width=0.32\textwidth]{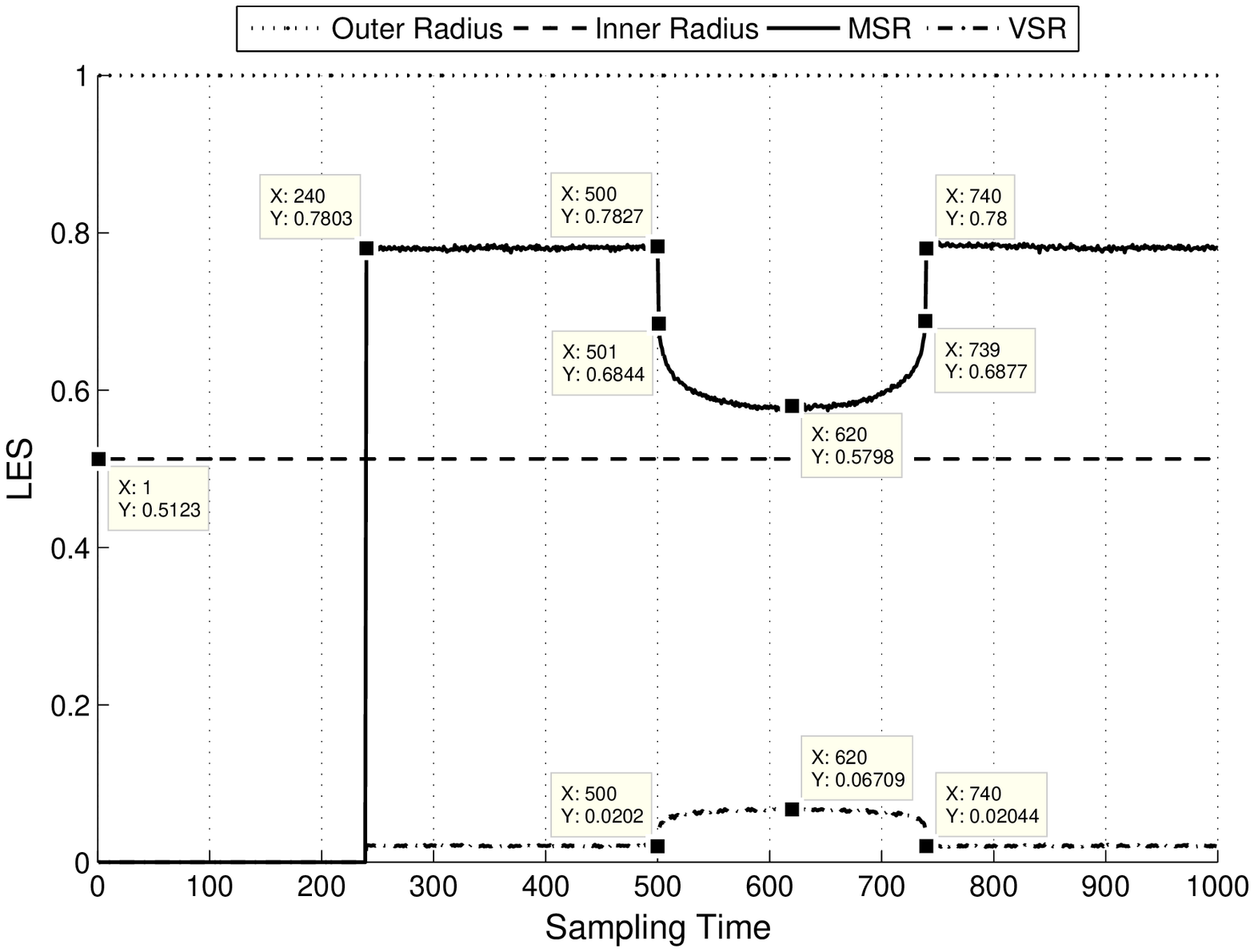}
\label{1-54}
}
\caption{$\kappa_\text{MSR}-t$ curves of standard matrix products in Case 1}
\label{1-MSR}

\end{figure*}

In case 1, assumed signals for each factor are shown in Tab.~\ref{Case-1}. The correlation analysis results are shown in Fig.~\ref{1-MSR}. It is noted that the correlation analysis begins at $t_s=240$, because the real-time split-window needs $T=240$ times of sampling data, including the present sampling and 239 times of historical sampling.

\begin{table}[H]
\centering
\caption{Assumed Signals for Each Factor in Case 1}
\label{Case-1}
\begin{tabular}{ccc}
\toprule[1pt]
\textbf{Bus} & \textbf{Sampling Time} & \textbf{Active Load(MW)} \\
\hline
\multirow{2}{*}{117} & $t_s=1 \sim 500$ & 20.0 \\
&$t_s=501 \sim 1000$ & 120.0 \\
\hline
Others & $t_s=1 \sim 1000$ &Unchanged \\
\hline
\end{tabular}
\end{table}

In Fig.~\ref{1-0}, based on the $\kappa_\text{MSR}\!-\!t$ curve, we can detect signals based on analyses below:\\
\small{}
I. During the sampling time $t_s \! = \! 240 \! \sim \! 500$, $\kappa_\text{MSR}$ of status matrix remains constant around 0.86, between the outer and inner circle; it means the system status is normal without signals.\\
II. At $t_s \! = \! 501$, $\kappa_\text{MSR}$ starts to decline (from 0.8638 to 0.7002), and deviates from the predicted ring (the inner radius is 0.7130); it means there exist signals that change the system status.\\
III. $\kappa_\text{MSR}$ increases back to 0.8662 at $t_s=740$ and remains constant inside the ring afterwards.\\

\normalsize{}
First, we can tell that the signals occur right at $t_s \! = \! 501$. Moreover, in our method, the impact of a signal to MSR will delay for a duration of $T$, due to historical sampling data in the real-time split-window. In Fig.~\ref{1-0}, the signal area (U-shaped curve) is $t_s\!=\!501\!\sim\!740$, so we can calculate the actual duration of signals as $740\!-\!501\!+\!1\!-\!T\!=\!0$. Therefore, we can speculate that there are instantaneous signals occurring at $t_s \! = \! 501$.

In order to find out the causes to above signals, we conduct correlation analysis for each factor. In Fig.~\ref{1-117}, when we augment load data of bus 117, $\kappa_\text{MSR}$ of the signal area ($t_s\!=\! 501 \! \sim \! 740$) decreases dramatically (from 0.7812 to 0.2998), below the inner circle (0.5123); it indicates strong correlations between the load of bus 117 and the system status. On the other hand, in Fig.~\ref{1-54}, $\kappa_\text{MSR}$ remains inside the ring throughout the signal area; it indicates poor correlations between the load of bus 54 and the system status.

\begin{figure}[!h]
\centering
\includegraphics[width=0.45\textwidth]{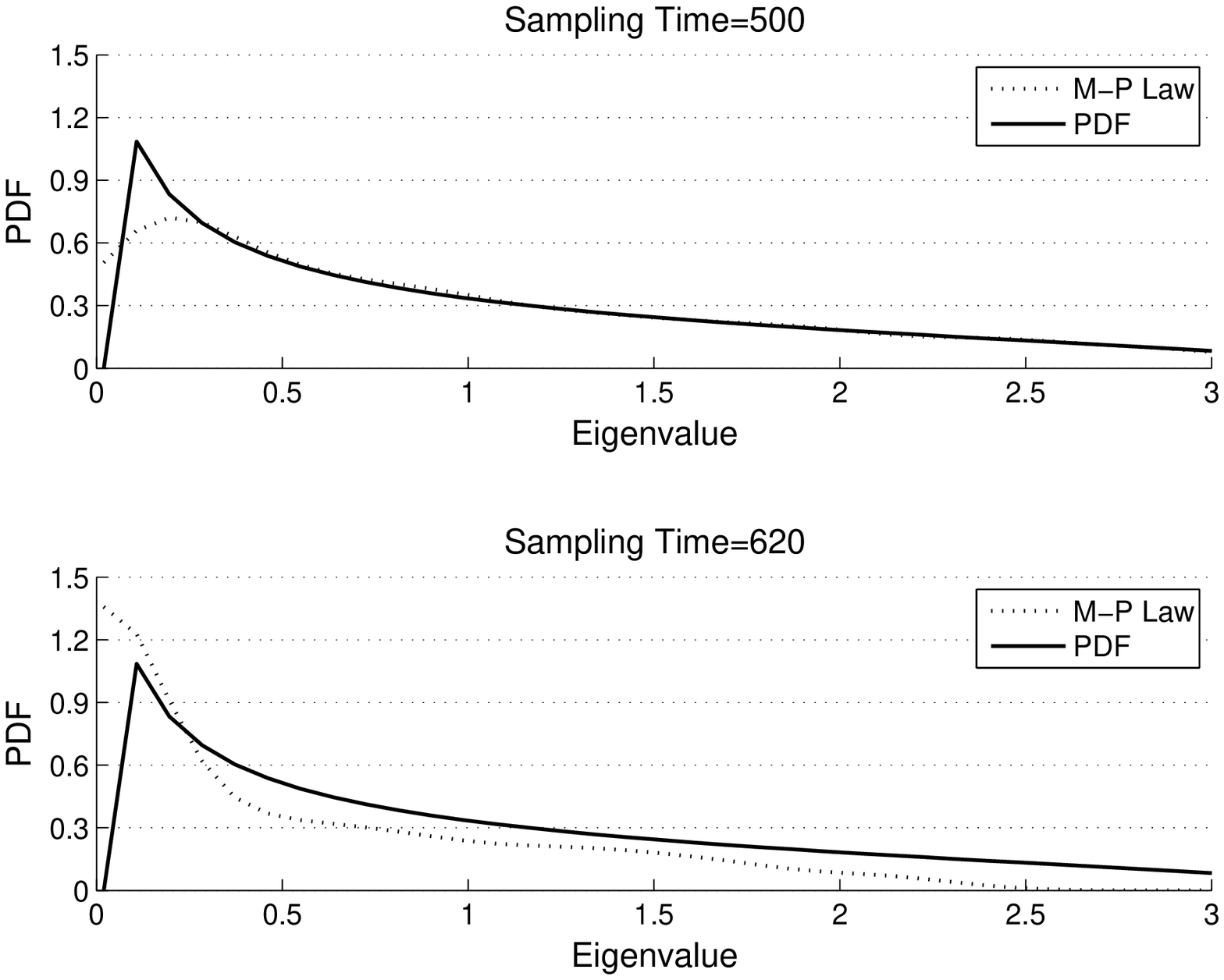}
\caption{$\kappa_\text{KDE}-\lambda$ curves of standard matrix products in case 1: the data source is the augmented matrix, including load data of bus 117}
\label{1-117-MP}

\centering
\includegraphics[width=0.45\textwidth]{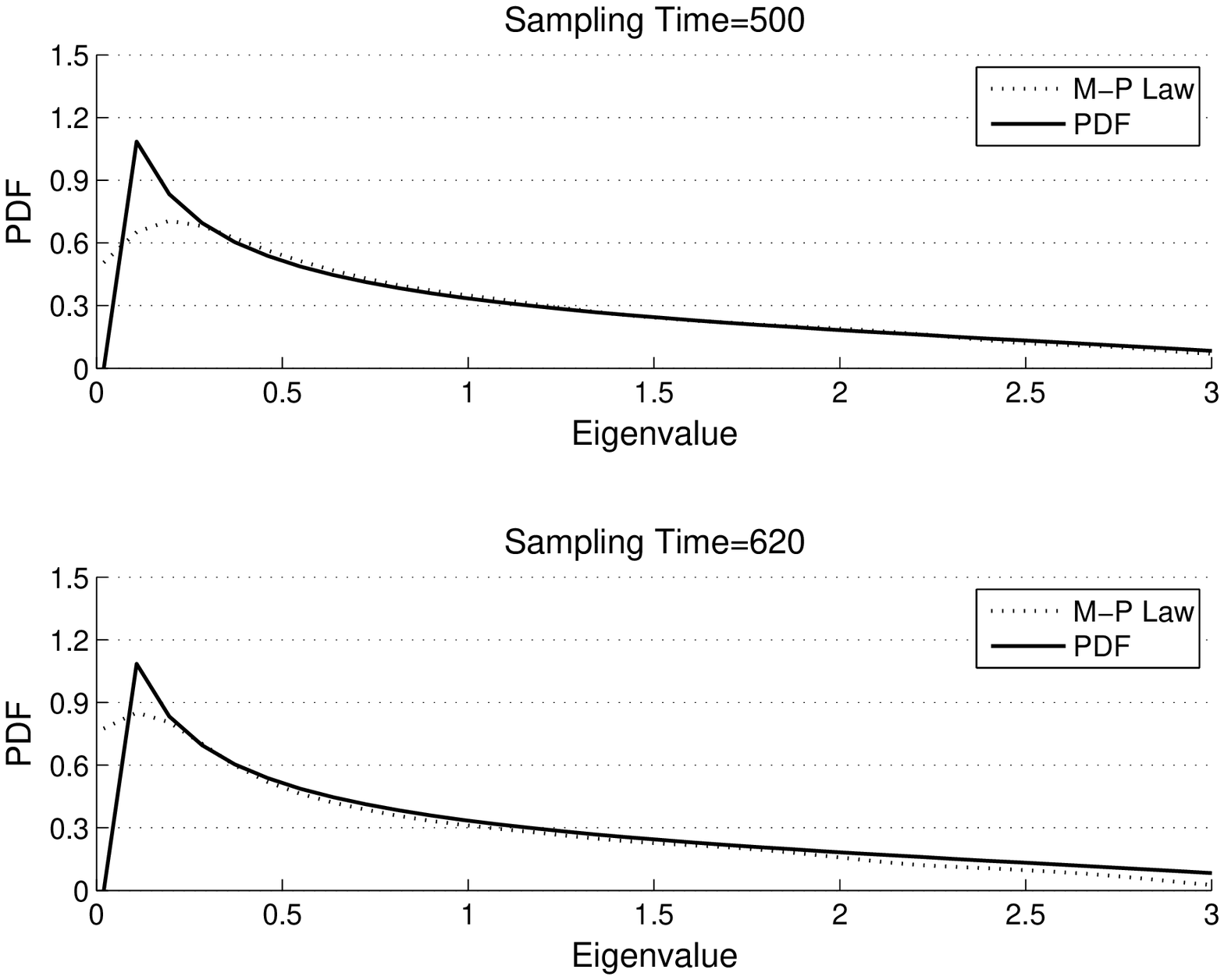}
\caption{$\kappa_\text{KDE}-\lambda$ curves of standard matrix products in case 1: the data source is the augmented matrix, including load data of bus 54}
\label{1-54-MP}
\end{figure}

Besides, we can also determine the correlations by $f_{\text{KDE}}-\lambda$ curves in Fig.~\ref{1-117-MP} and Fig.~\ref{1-54-MP}. In Fig.~\ref{1-117-MP}, at $t_s=620$ (inside the signal area), when we augment load data of bus 117, $f_{\text{KDE}}$ deviates from $f_{\text{MP}}$; it indicates strong correlations between the load of bus 117 and the system status. However, in Fig.~\ref{1-54-MP}, $f_{\text{KDE}}$ with the load of bus 54 accords with $f_{\text{MP}}$ at $t_s=620$; it indicates poor correlations between the load of bus 54 and the system status.

As a result, we deduce that the load of bus 117, but not bus 54, is the cause for instantaneous signals at $t_s \! = \! 501$. This analysis result accords with assumed signals in Tab.~\ref{Case-1}. In fact, we only add signals to the active load of bus 117. Specifically, the active load of bus 117 increases from 20 MW to 120 MW right at $t_s \! = \! 501$.

\begin{figure}[!h]
\centering
\subfloat[$t_s=500$]{
\includegraphics[width=0.22\textwidth]{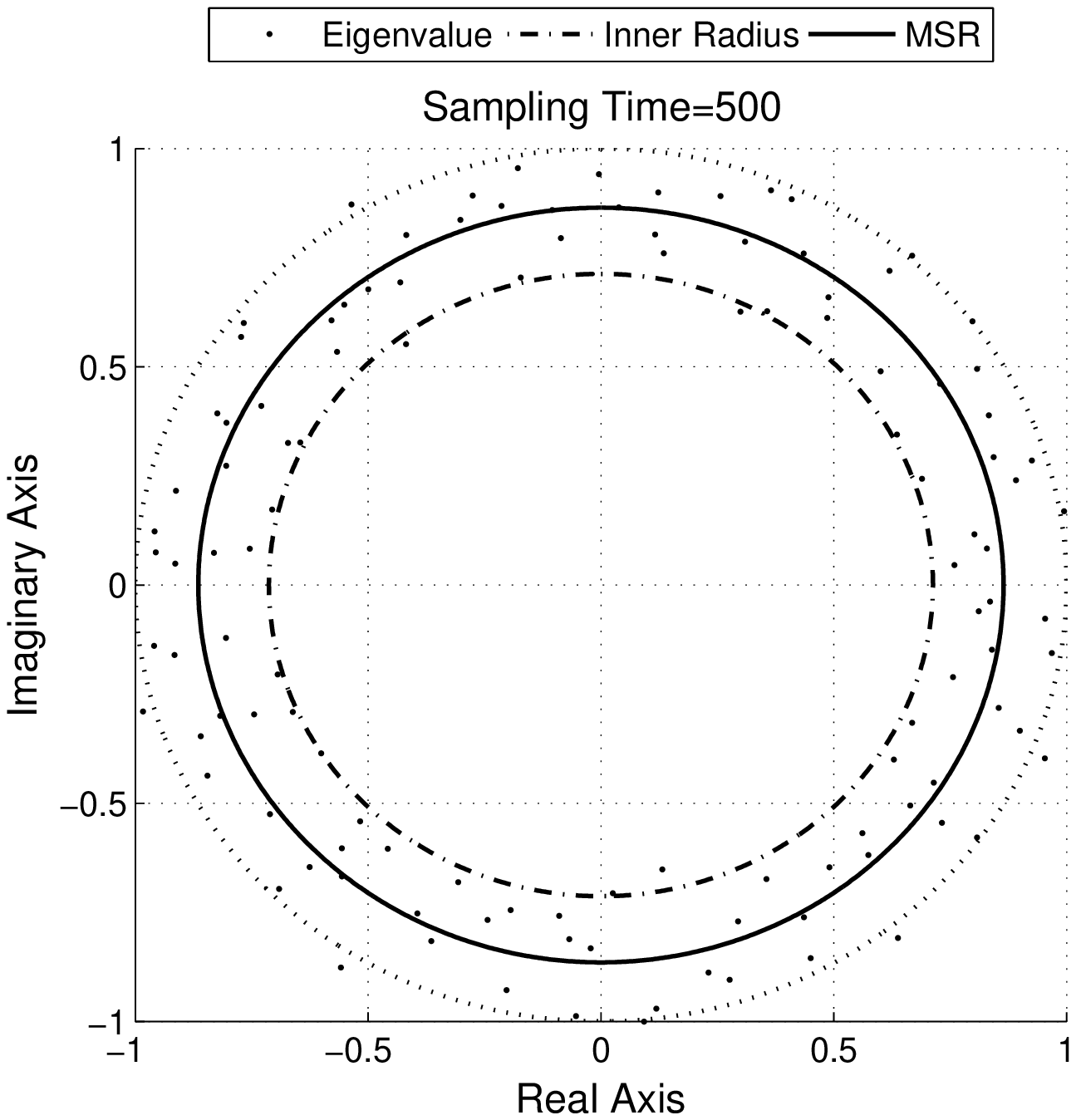}
\label{1-0-RL-500}
}
\subfloat[$t_s=620$]{
\includegraphics[width=0.22\textwidth]{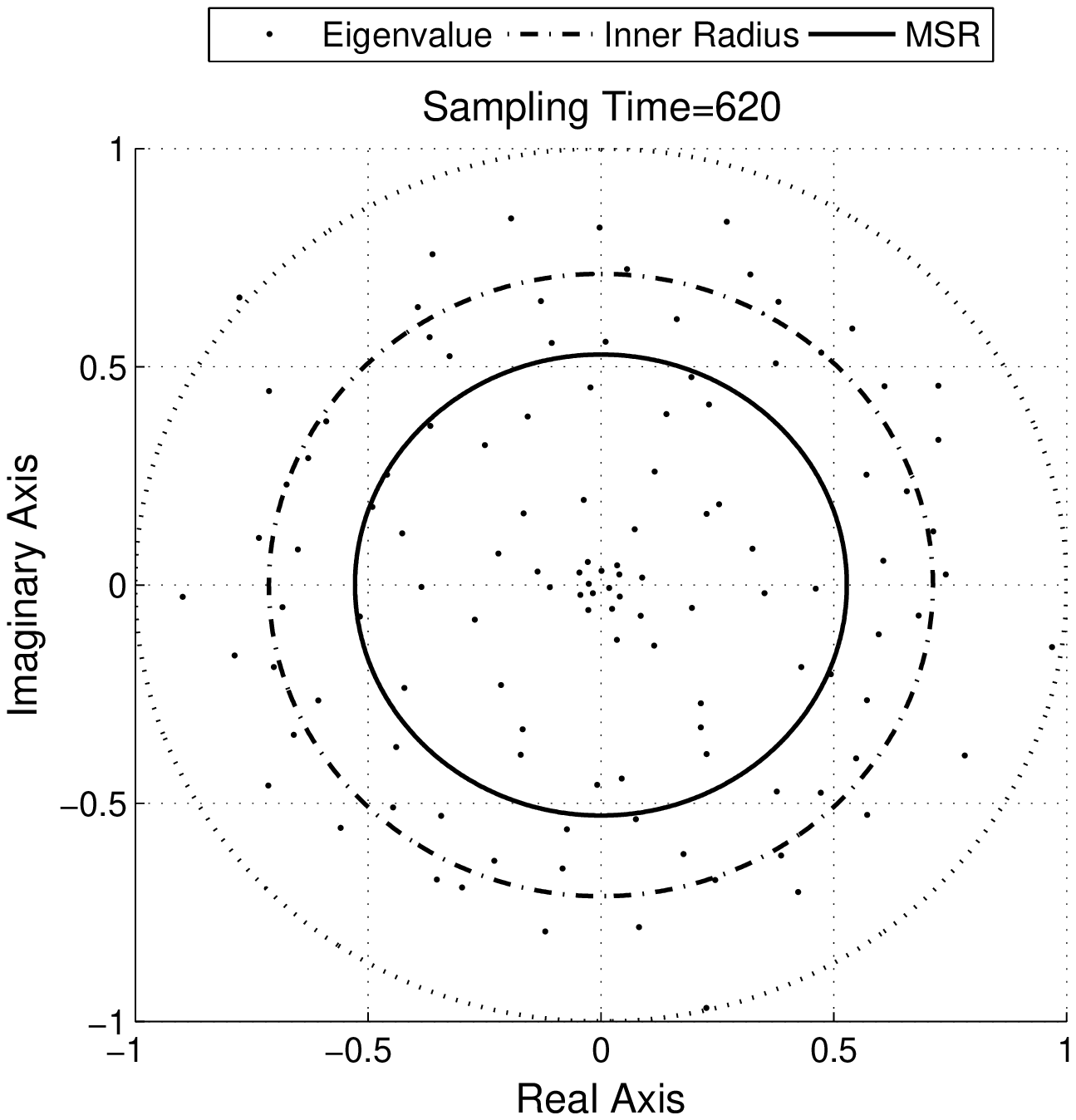}
\label{1-0-RL-620}
}
\caption{Eigenvalue distributions of standard matrix products in case 1: the data source is the status matrix.}
\label{1-0-RL}

\centering
\subfloat[$t_s=500$]{
\includegraphics[width=0.22\textwidth]{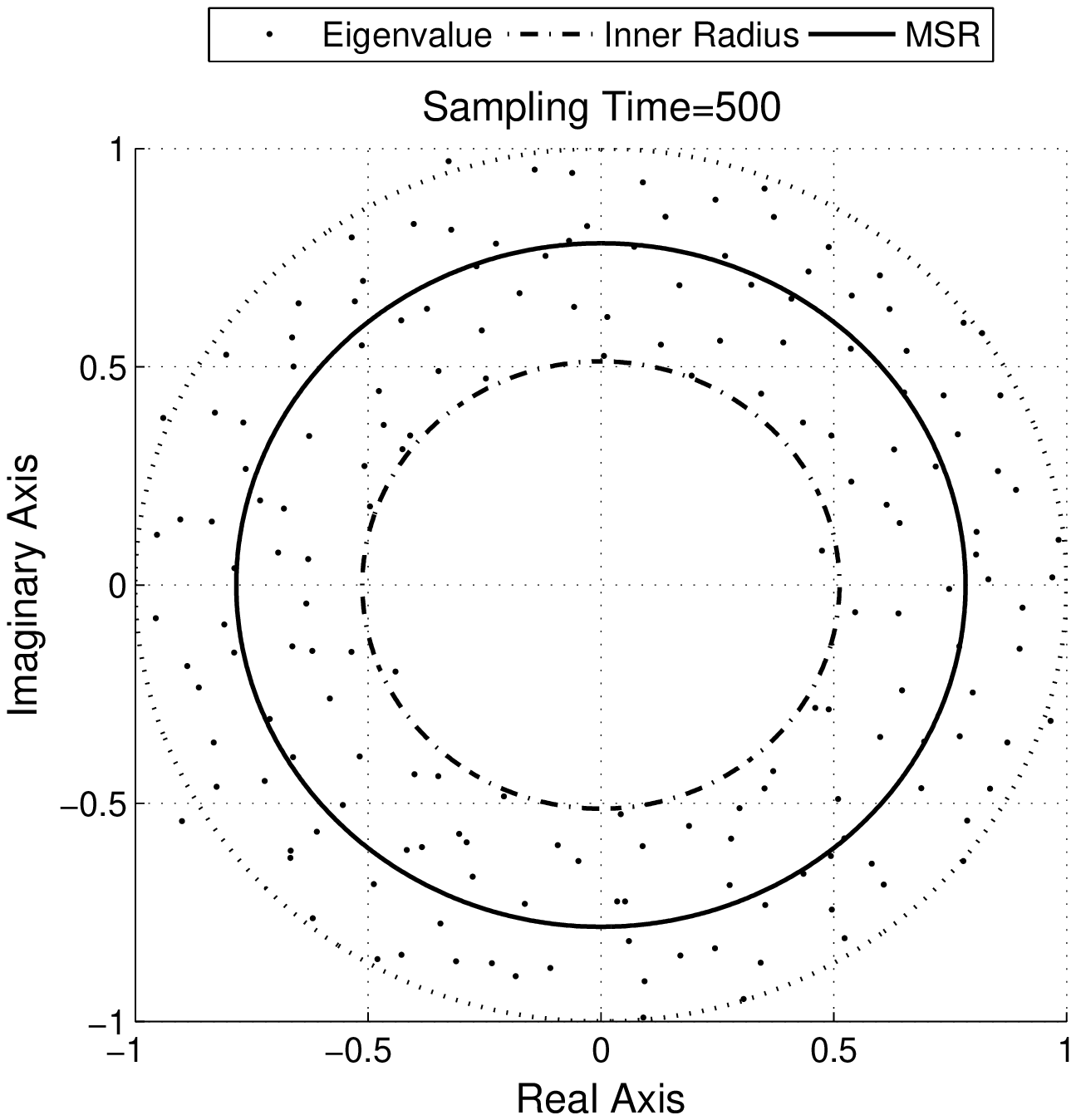}
\label{1-117-RL-500}
}
\subfloat[$t_s=620$]{
\includegraphics[width=0.22\textwidth]{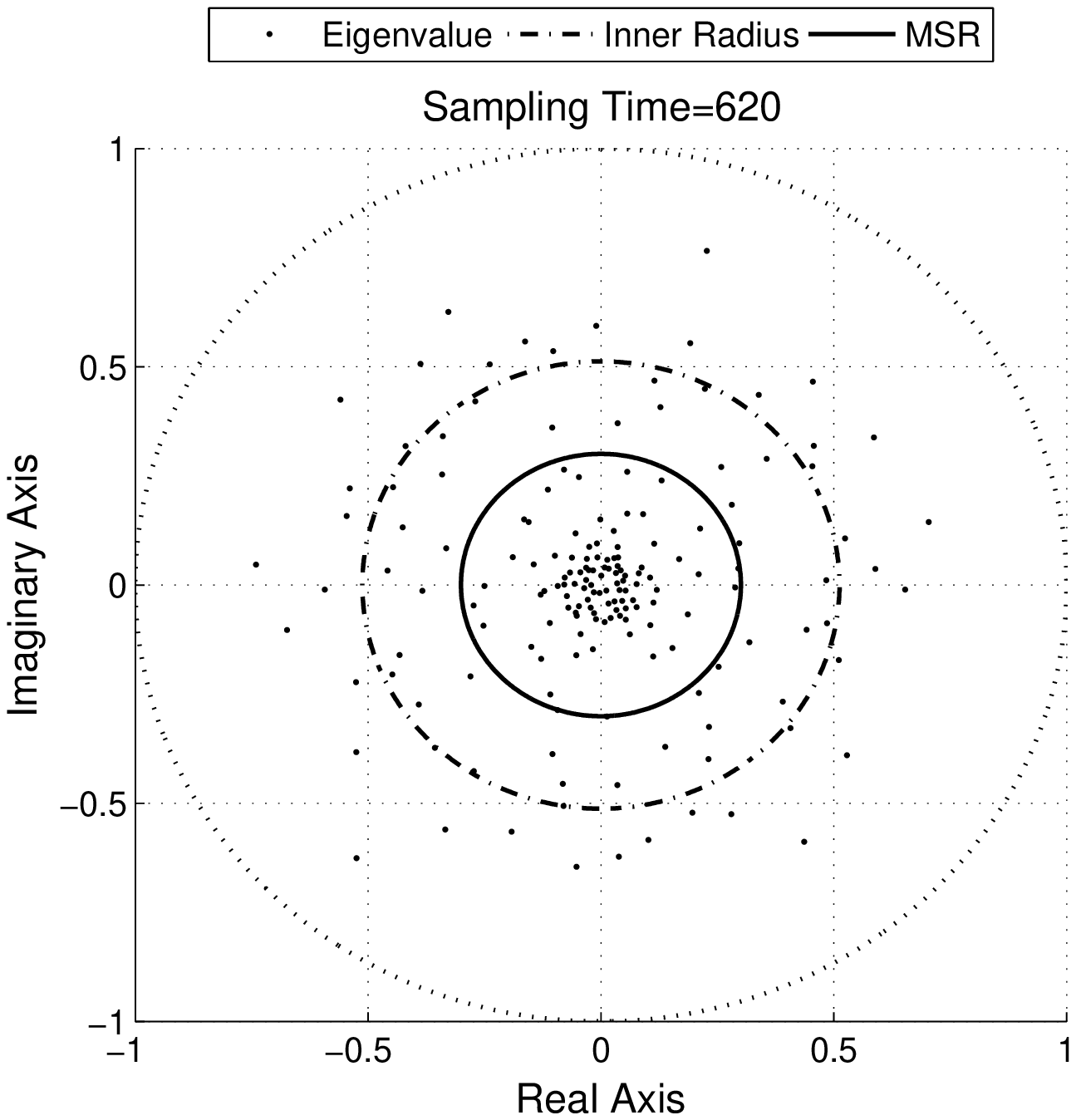}
\label{1-117-RL-620}
}
\caption{Eigenvalue distributions of standard matrix products in case 1: the data source is the augmented matrix, including load data of bus 117.}
\label{1-117-RL}

\centering
\subfloat[$t_s=500$]{
\includegraphics[width=0.22\textwidth]{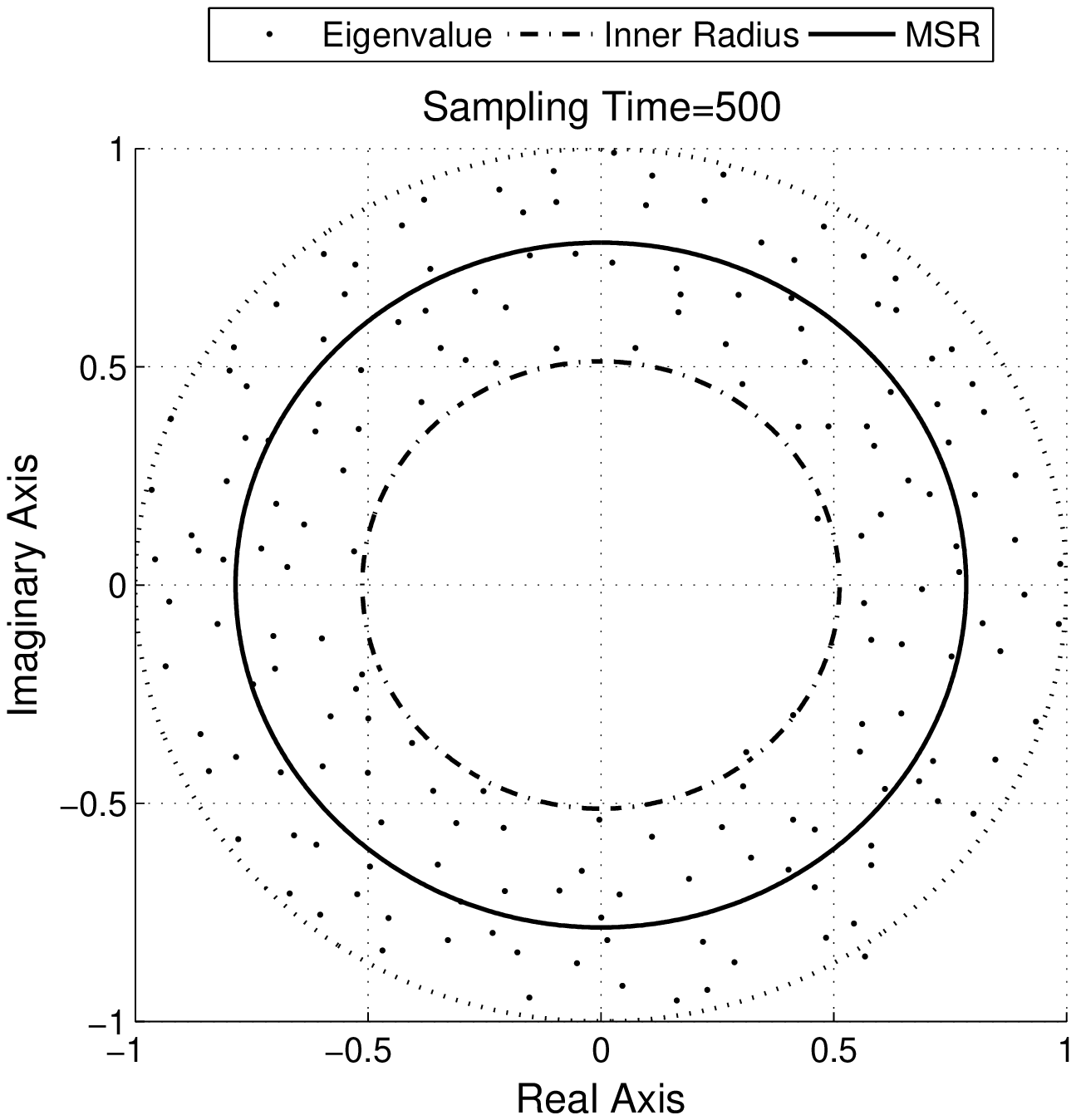}
\label{1-54-RL-500}
}
\subfloat[$t_s=620$]{
\includegraphics[width=0.22\textwidth]{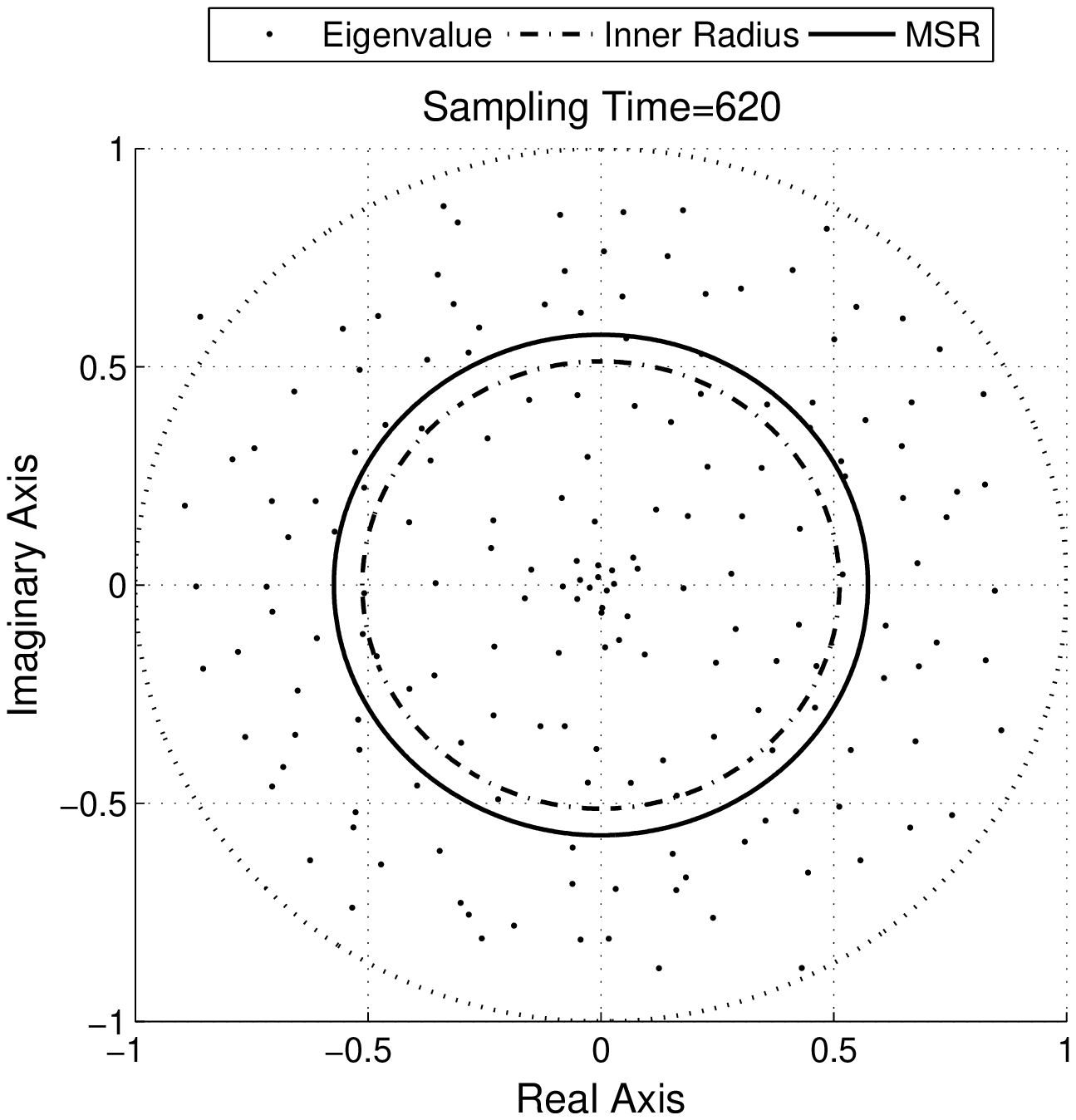}
\label{1-54-RL-620}
}
\caption{Eigenvalue distributions of standard matrix products in case 1: the data source is the augmented matrix, including load data of bus 54.}
\label{1-54-RL}
\end{figure}

\subsubsection{Case 2 \!-\! Peak and dip signals in the load of bus 117}

\

In case 2, assumed signals for each factor are shown in Tab.~\ref{Case-2}. The correlation analysis results are shown in Fig.~\ref{2-MSR}.

\begin{table}[H]
\centering
\caption{Assumed Signals for Each Factor in Case 2}
\label{Case-2}
\begin{tabular}{ccc}
\toprule[1pt]
\textbf{Bus} & \textbf{Sampling Time} & \textbf{Active Load(MW)} \\
\hline
\multirow{5}{*}{117} & $t_s=1 \sim 300$ & 60.0 \\
&$t_s=301 \sim 350$ & 120.0 \\
&$t_s=351 \sim 650$ & 60.0 \\
&$t_s=651 \sim 700$ & 20.0 \\
&$t_s=701 \sim 1000$ & 60.0 \\
\hline
Others & $t_s=1 \sim 1000$ &Unchanged \\
\hline
\end{tabular}
\end{table}

In Fig.~\ref{2-0}, based on the $\kappa_\text{MSR}-t$ curve, we can detect signals below:

I. From $t_s\!=\!301$ to $t_s\!=\!590$, the $\kappa_\text{MSR}-t$ curve is U-shaped beyond the predict ring. It indicates that the signals occur at $t_s\!=\!301$, and the signal area is $t_s\!=\!301\sim\!590$.

II. From $t_s\!=\!650$ to $t_s\!=\!940$, the $\kappa_\text{MSR}-t$ curve is U-shaped beyond the predict ring. It indicates that the signals occur at $t_s\!=\!650$, and the signal area is $t_s\!=\!650\sim\!940$.

In consideration of the delayed effect of a signal to MSR, we can calculate actual durations of above signals as $590\!-\!301\!+\!1\!-\!T\!=\!50$ and $940\!-\!651\!+\!1\!-\!T\!=\!50$. Therefore, we can speculate that there are continuous signals occurring at $t_s\!=\!301\!\sim\!351$ and $t_s\!=\!651\!\sim\!701$.

In Fig.~\ref{2-117}, when we augment load data of bus 117, $\kappa_\text{MSR}$ inside two signal areas declines remarkably and deviates from the predicted ring (from 0.7858 to 0.3538 and from 0.7830 to 0.3846); it indicates strong correlations between the load of bus 117 and the system status. On the other hand, in Fig.~\ref{2-54}, when we augment load data of bus 54, $\kappa_\text{MSR}$ inside both signal areas remains in the predicted ring throughout the signal area; it indicates poor correlations between the load of bus 54 and the system status. As a result, we deduce that the load of bus 117, but not bus 54, is the cause for continuous signals during $t_s\!=\!301\!\sim\!351$ and $t_s\!=\!651\!\sim\!701$.

Above analyses accord with assumed signals in Tab.~\ref{Case-2}. In this case, we only add signals to the active load of bus 117. To be specific, the active load of bus 117 has a peak during $t_s\!=\!301\!\sim\!351$, and a dip during $t_s\!=\!651\!\sim\!701$.

\subsubsection{Case 3 \!-\! gradual signals in the load of bus 54}

\

In case 3, assumed signals for each factor are shown in Tab.~\ref{Case-3}. The correlation analysis results are shown in Fig.~\ref{3-MSR}.

\begin{table}[H]
\centering
\caption{Assumed Signals for Each Factor in Case 3}
\label{Case-3}
\begin{tabular}{ccc}
\toprule[1pt]
\textbf{Bus} & \textbf{Sampling Time} & \textbf{Active Load(MW)} \\
\hline
\multirow{10}{*}{54} & $t_s=1 \sim 300$ & 113.0 \\
&$t_s=301 \sim 350$ & 135.6 \\
&$t_s=351 \sim 400$ & 158.2 \\
&$t_s=401 \sim 450$ & 180.8 \\
&$t_s=451 \sim 500$ & 203.4 \\
&$t_s=501 \sim 550$ & 226.0 \\
&$t_s=551 \sim 600$ & 248.6 \\
&$t_s=601 \sim 650$ & 271.2 \\
&$t_s=651 \sim 700$ & 293.8 \\
&$t_s=701 \sim 1000$ & 316.4 \\
\hline
Others & $t_s=1 \sim 1000$ &Unchanged \\
\hline
\end{tabular}
\end{table}

In Fig.~\ref{3-0}, the U-shaped curve is from $t_s\!=\!301$ to $t_s\!=\!940$. It indicates signals occurring at $t_s\!=\!301$. In consideration of the delayed effect of a signal to MSR, we can calculate actual durations of above signals as $940\!-\!301\!+\!1\!-\!T\!=\!400$. Therefore, we can speculate that there are continuous signals occurring at $t_s\!=\!301\!\sim\!701$.

In Fig.~\ref{3-54}, when we augment load data of bus 54, $\kappa_\text{MSR}$ inside the signal area declines remarkably and deviates from the predicted ring (from 0.7803 to 0.4227); it indicates strong correlations between the load of bus 54 and the system status. On the other hand, in Fig.~\ref{3-117}, when we augment load data of bus 117, $\kappa_\text{MSR}$ remains in the predicted ring throughout the signal area; it indicates poor correlations between the load of bus 117 and the system status. As a result, we deduce that the load of bus 54, but not bus 117, is the cause for continuous signals during $t_s\!=\! 301 \! \sim \! 701$.

Above analyses is in accordance with assumed signals in Tab.~\ref{Case-3}. In this case, we only add signals to the active load of bus 54. In detail, the active load of bus 54 increase gradually during $t_s\!=\!301\!\sim\!701$.

\subsection{Correlation Analysis for Multiple Factors}
\begin{figure*}[ht]
\centering
\subfloat[Data source: status matrix]{
\includegraphics[width=0.32\textwidth]{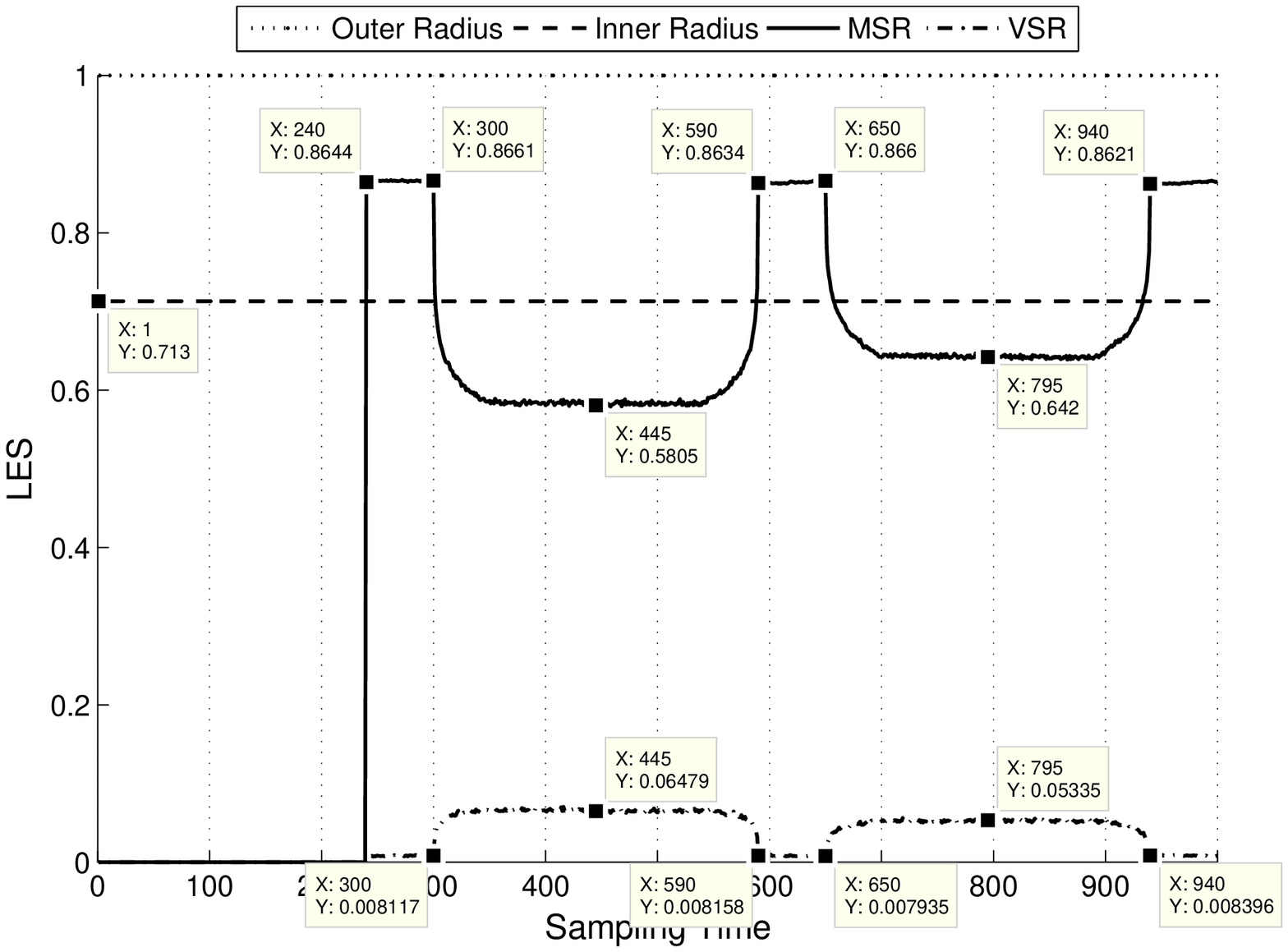}
\label{2-0}
}
\subfloat[Data source: augmented matrix including load data of bus 117]{
\includegraphics[width=0.32\textwidth]{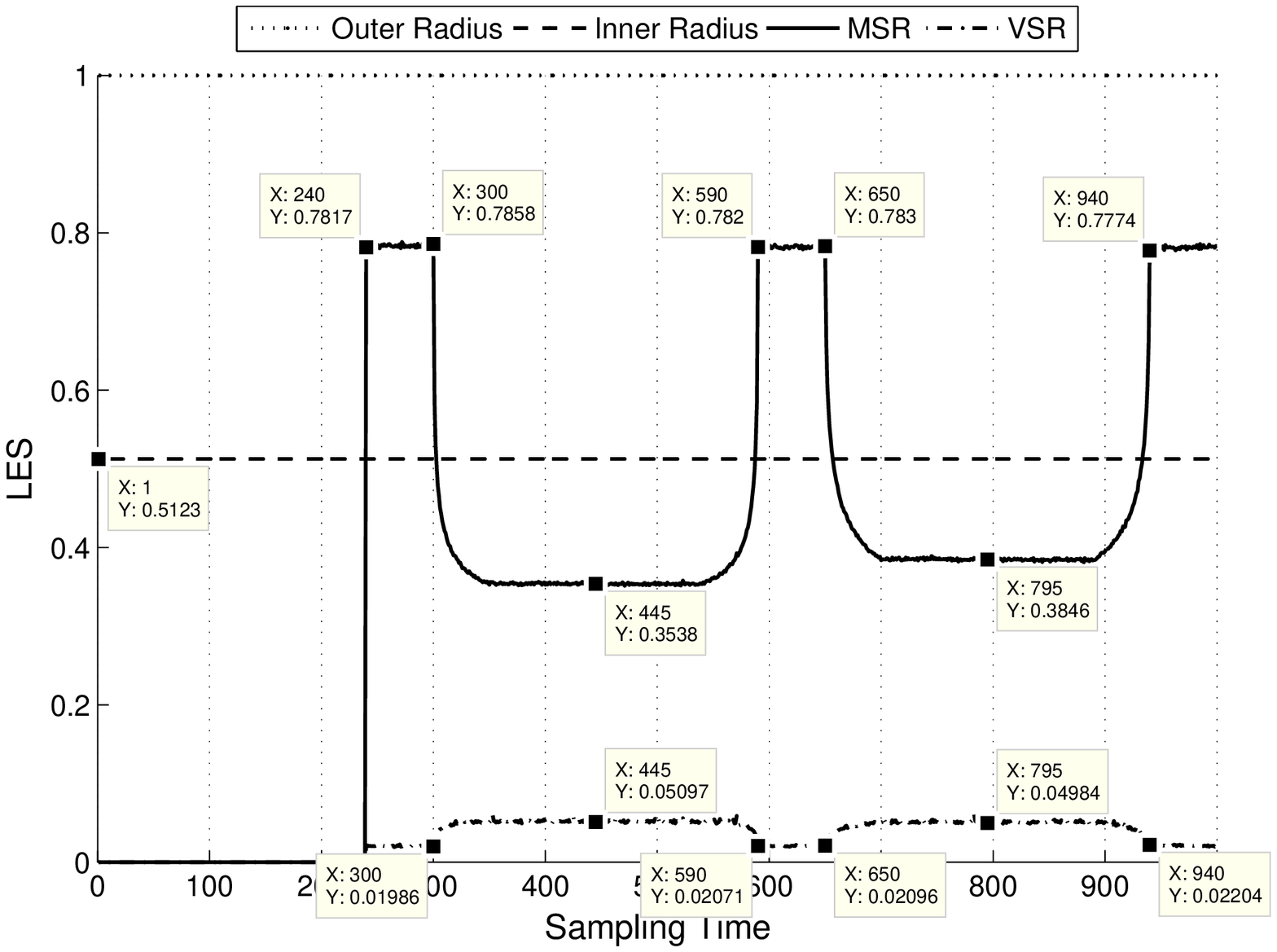}
\label{2-117}
}
\subfloat[Data source: augmented matrix including load data of bus 54]{
\includegraphics[width=0.32\textwidth]{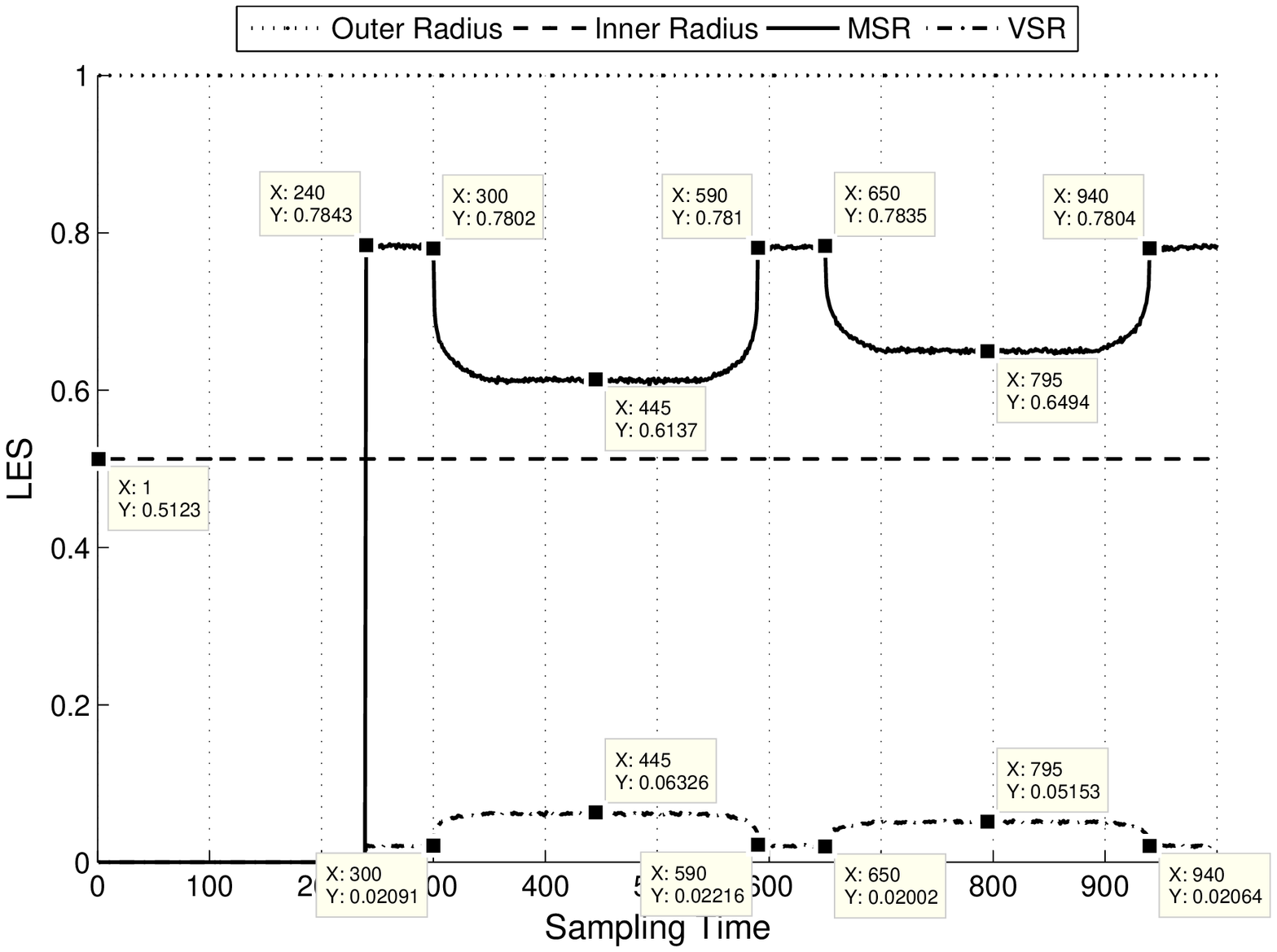}
\label{2-54}
}
\caption{$\kappa_\text{MSR}-t$ curves of standard matrix products in Case 2}
\label{2-MSR}

\centering
\subfloat[Data source: status matrix]{
\includegraphics[width=0.32\textwidth]{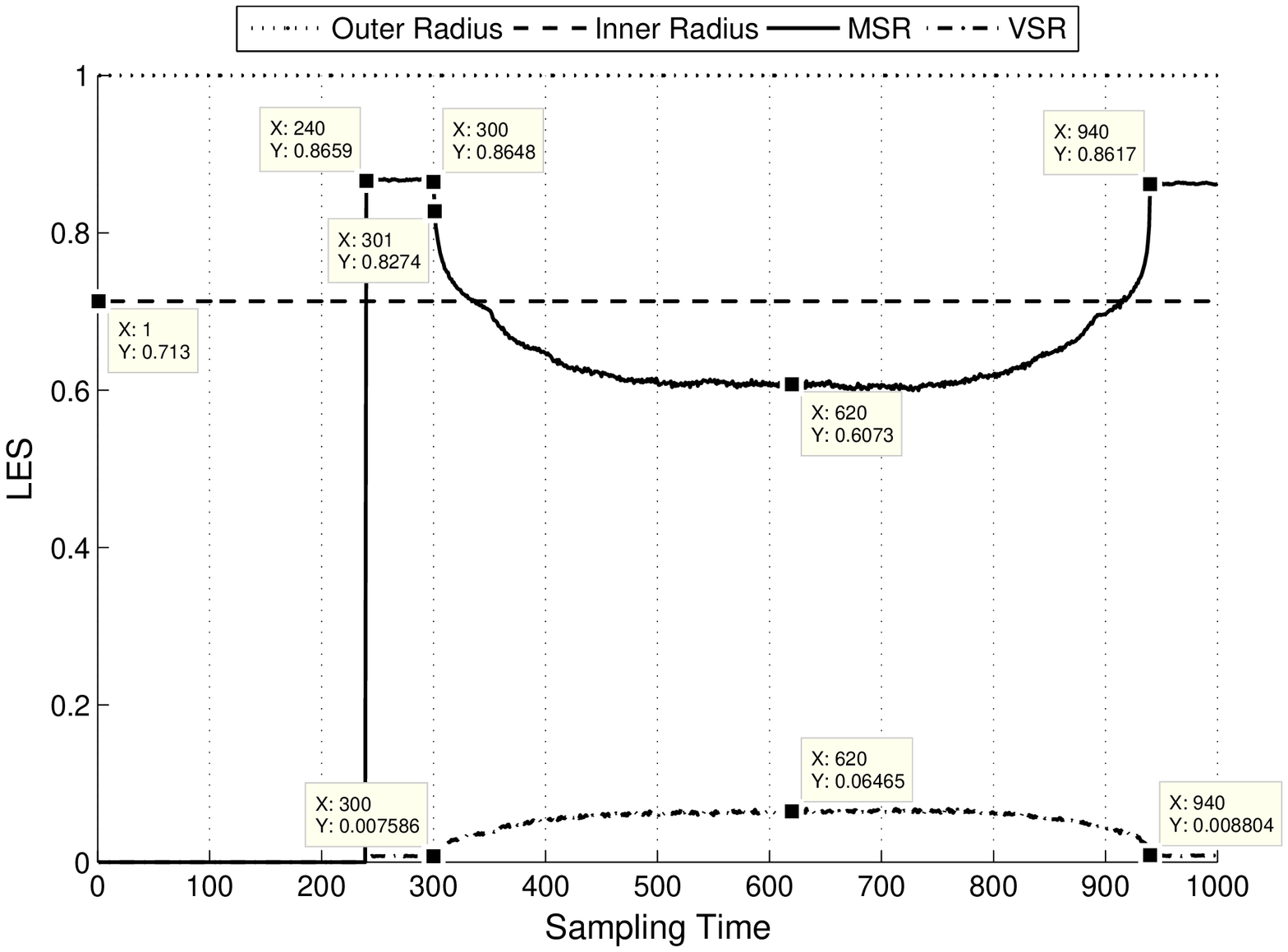}
\label{3-0}
}
\subfloat[Data source: augmented matrix including load data of bus 117]{
\includegraphics[width=0.32\textwidth]{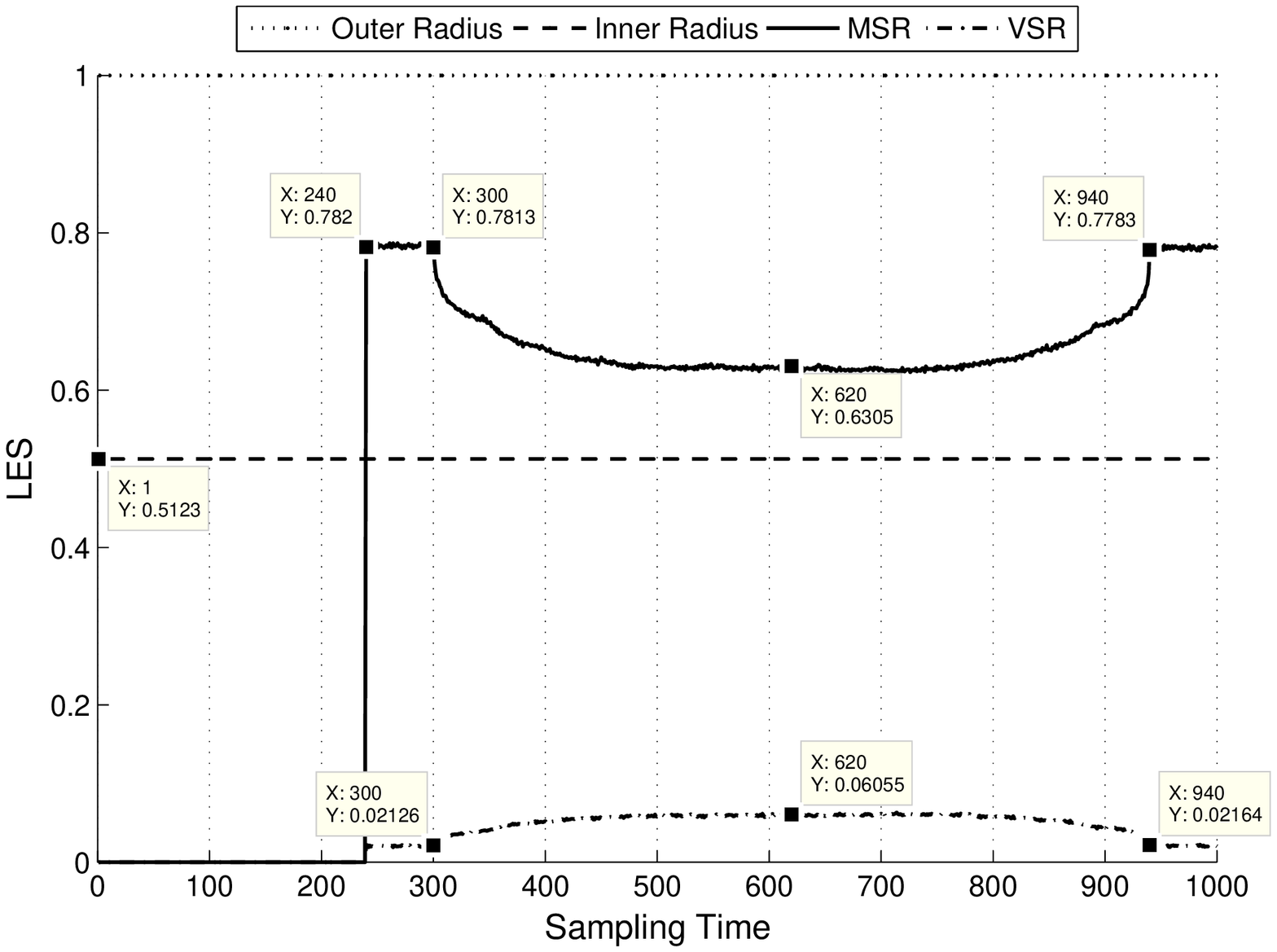}
\label{3-117}
}
\subfloat[Data source: augmented matrix including load data of bus 54]{
\includegraphics[width=0.32\textwidth]{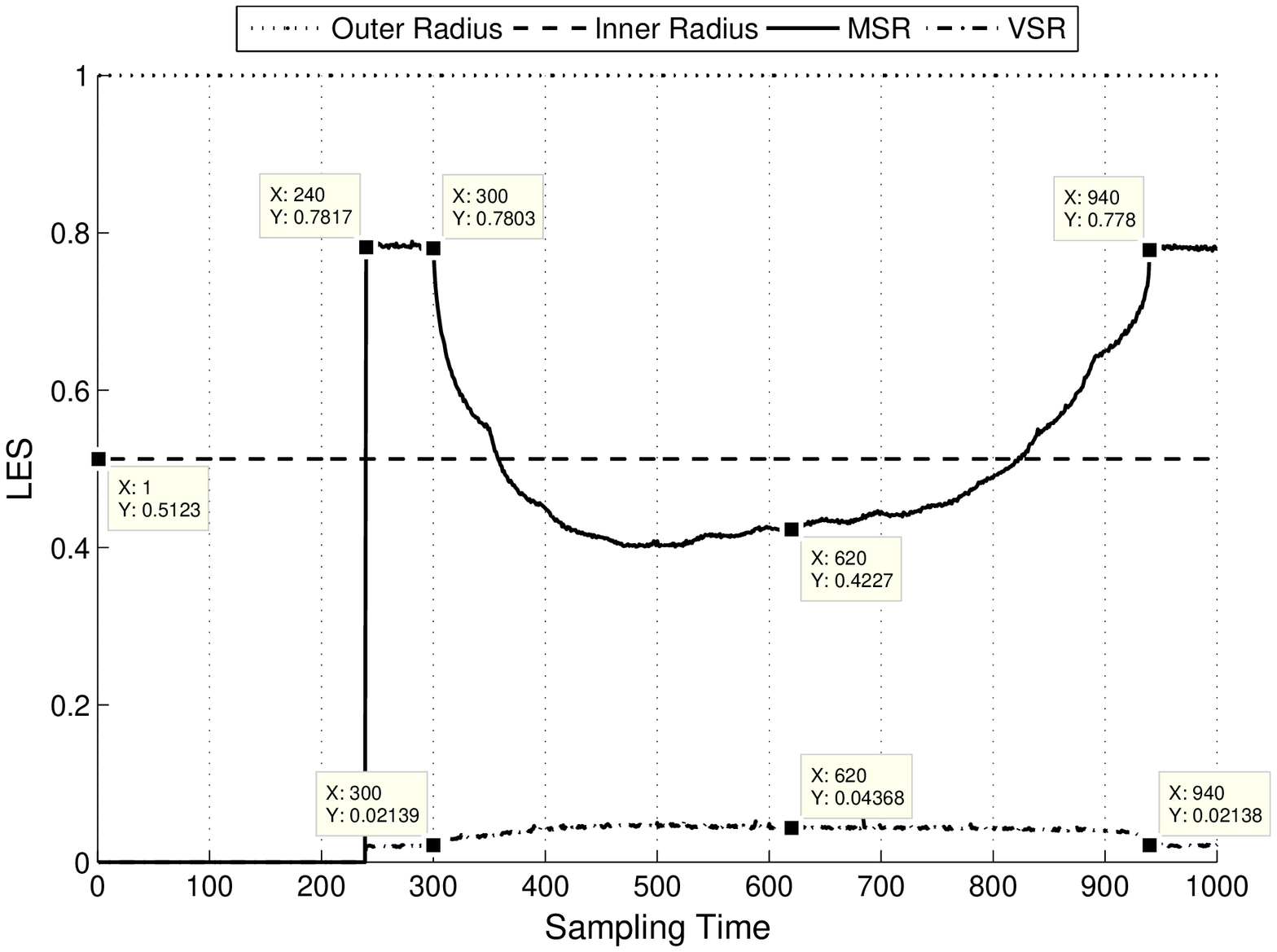}
\label{3-54}
}
\caption{$\kappa_\text{MSR}-t$ curves of standard matrix products in Case 3}
\label{3-MSR}

\centering
\subfloat[Data source: status matrix]{
\includegraphics[width=0.32\textwidth]{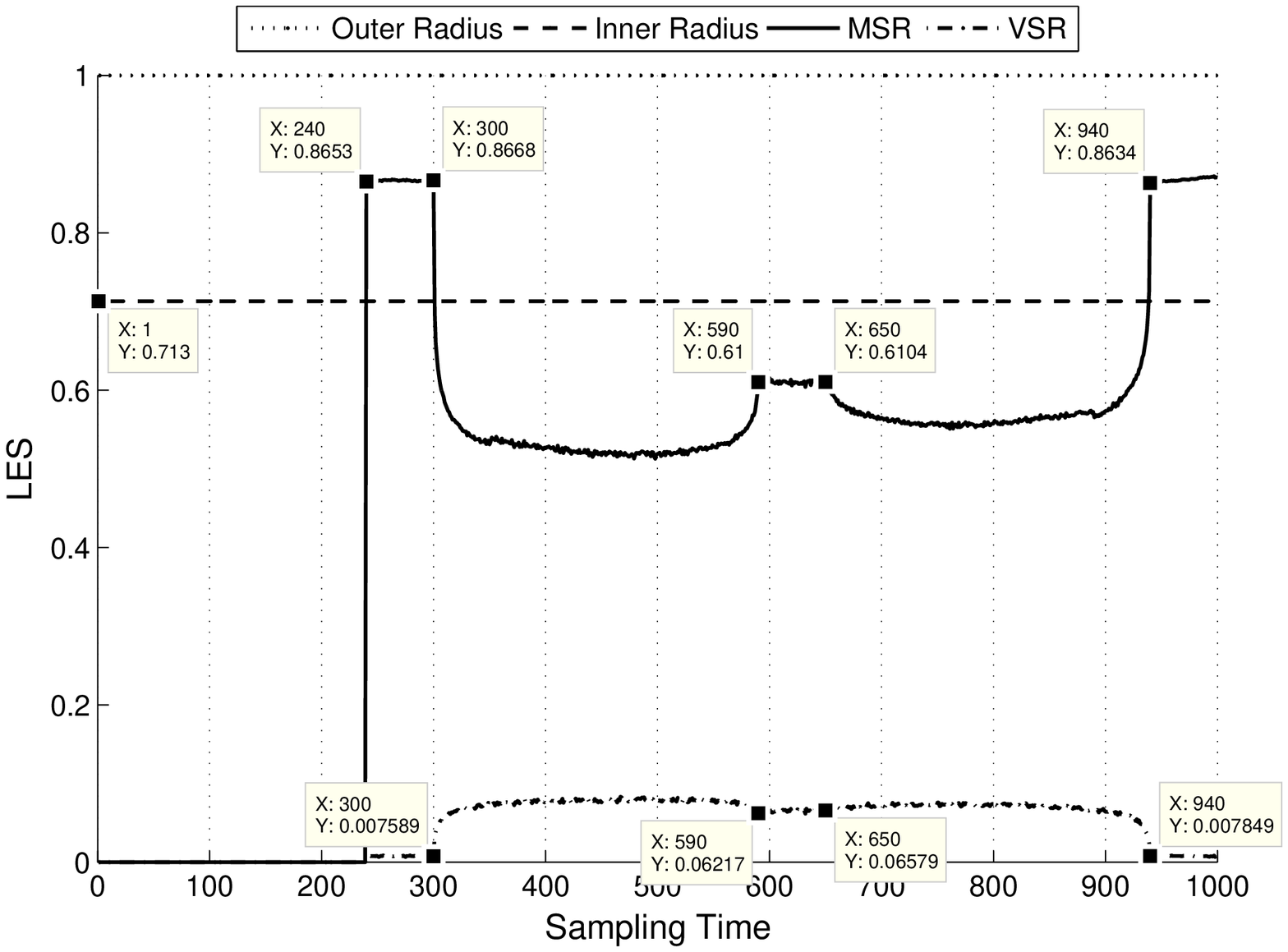}
\label{4-0}
}
\subfloat[Data source: augmented matrix including load data of bus 117]{
\includegraphics[width=0.32\textwidth]{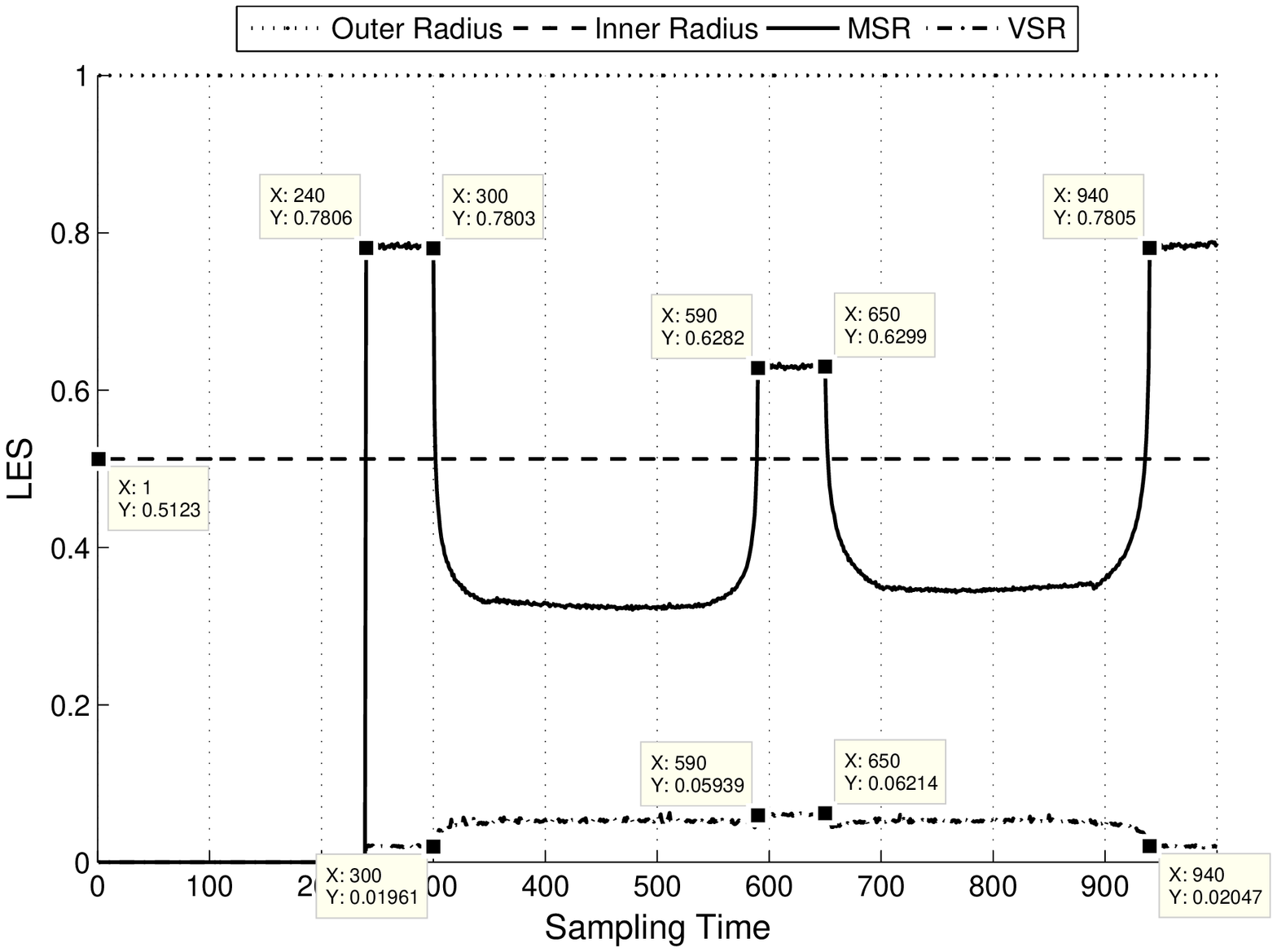}
\label{4-117}
}
\subfloat[Data source: augmented matrix including load data of bus 54]{
\includegraphics[width=0.32\textwidth]{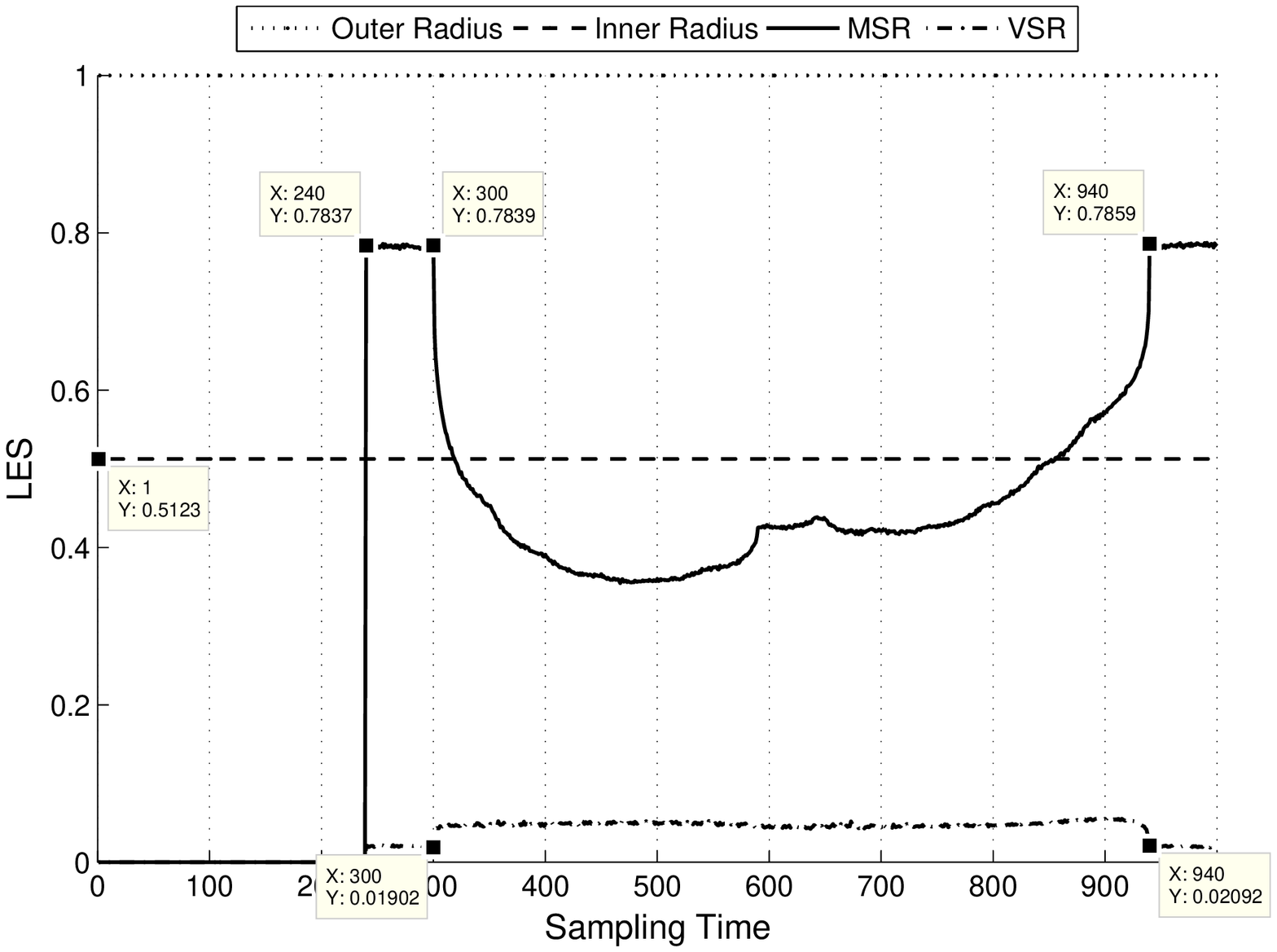}
\label{4-54}
}
\caption{$\kappa_\text{MSR}-t$ curves of standard matrix products in Case 4}
\label{4-MSR}

\end{figure*}

In case 4, assumed signals for each factor are shown in Tab.~\ref{Case-4}. The correlation analysis results are shown in Fig.~\ref{4-MSR}.

\begin{table}[H]
\centering
\caption{Assumed Signals for Each Factor in Case 4}
\label{Case-4}
\begin{tabular}{ccc}
\toprule[1pt]
\textbf{Bus} & \textbf{Sampling Time} & \textbf{Active Load(MW)} \\
\hline
\multirow{5}{*}{117} & $t_s=1 \sim 300$ & 60.0 \\
&$t_s=301 \sim 350$ & 120.0 \\
&$t_s=351 \sim 650$ & 60.0 \\
&$t_s=651 \sim 700$ & 20.0 \\
&$t_s=701 \sim 1000$ & 60.0 \\
\hline
\multirow{10}{*}{54} & $t_s=1 \sim 300$ & 113.0 \\
&$t_s=301 \sim 350$ & 135.6 \\
&$t_s=351 \sim 400$ & 158.2 \\
&$t_s=401 \sim 450$ & 180.8 \\
&$t_s=451 \sim 500$ & 203.4 \\
&$t_s=501 \sim 550$ & 226.0 \\
&$t_s=551 \sim 600$ & 248.6 \\
&$t_s=601 \sim 650$ & 271.2 \\
&$t_s=651 \sim 700$ & 293.8 \\
&$t_s=701 \sim 1000$ & 316.4 \\
\hline
Others & $t_s=1 \sim 1000$ &Unchanged \\
\hline
\end{tabular}
\end{table}

Based on the $\kappa_\text{MSR}-t$ curve in Fig.~\ref{4-0}, two kinds of signals are detected:

I. Two U-shaped curves are found during $t_s\!=\!301\!\sim\!590$ and $t_s\!=\!651\!\sim\!940$. Referring to the analysis in case 2, it indicates two continous signals during $t_s\!=\!301\!\sim\!351$ and $t_s\!=\!651\!\sim\!701$.

I. The Third curve are found during $t_s\!=\!301\!\sim\!940$. Referring to the analysis in case 3, it indicates continuous signals during $t_s\!=\!301\!\sim\!700$.

During the first two signal areas ($t_s\!=\!301\!\sim\!590$ and $t_s\!=\!651\!\sim\!940$), when we augment the load of bus 117, $\kappa_\text{MSR}$ deviates from Ring Law, shown in Fig.~\ref{4-117}. During the third signal area ($t_s\!=\!301\!\sim\!940$), when we augment the load of bus 54, the deviation of $\kappa_\text{MSR}$ is shown in Fig.~\ref{4-54}. As a result, we can achieve following speculations:

I. The load of bus 117 affects the system status during $t_s\!=\!301\!\sim\!351$ and $t_s\!=\!651\!\sim\!701$.

II. The load of bus 54 affects the system status during $t_s\!=\!301\!\sim\!700$.

These speculations are validated by Table.~\ref{Case-4}.

\subsection{More Discussions in Details}
Through above four cases, the effectiveness and performance of the correlation analysis method is verified under different scenarios. However, there are still some interesting details left.

First, we can observe that $\kappa_\text{MSR}\!-\!t$ curves in case 4 are approximately superposed by corresponding curves in case 2 and case 3. In fact, the assumed signals in case 4 are combined with those in case 2 and case 3. In conclusion, there should be some relationships between phenomena in power systems, and LES in mathematics.

Secondly, when data become more correlated, the VSR increases as well. This phenomenon can be explained based on the eigenvalue distributions of standard matrix products. In Fig.~\ref{1-0-RL-500}, Fig.~\ref{1-117-RL-500} and Fig.~\ref{1-54-RL-500}, at $t_s\!=\!500$, the eigenvalues distribute between the outer circle and the inner circle, conforming to the Ring Law; it indicates poor correlations in data. During the signal area (such as $t_s\!=\!620$), when we augment data of correlated factors, all the eigenvalues gather towards the circle center, shown in Fig.~\ref{1-117-RL-620}. When we augment data of irrelevant factors, only some of the eigenvalues gather towards the circle center, shown in Fig.~\ref{1-54-RL-620}.

Thirdly, to represent random fluctuations in the system and measuring errors in sampling data, white noise is introduced into both status data and factor data throughout the simulation. Observations in case studies indicate that $\kappa_\text{MSR}$ does not change dramatically in a system dominated by white noise. Therefore, $\kappa_\text{MSR}$ is a reliable indicator to identify signals from random fluctuations and measuring errors.

\section{Conclusion}
This paper proposes a data-driven method to reveal the correlations between factors and the power system status. First, for each factor, we construct an augmented matrix as a data source, combining status data and factors data reasonably. Secondly, to conduct real-time analysis, we use a specific split-window to obtain the raw data matrix at each sampling time. Then, we conduct correlation analysis for the raw data matrix based on random matrix theory (RMT). The mean spectral radius (MSR), a kind of linear eigenvalue statistics (LESs), is calculated as a correlation indicator; the kernel density estimation (KDE) is used as a assisted tool to reveal data correlations. The proposed method is direct, robust against bad data and universal to varieties of factors. Case studies demonstrate the effectiveness of the method for both single factor and multiple factors.

The current work is only a preliminary exploration of correlation analysis based on RMT. Much more researches are needed along this direction. The degree of correlations needs to be further quantified. For example, in a study period, we can use the integration of MSR to quantify the correlations. Besides, we can use data of subarea to construct the basic matrix; in this way we can fix the data missing problems. Furthermore, the proposed method can be used to reveal the correlation between any two types of variables, as long as combining their data reasonably in the data source. Our method can be applied for specific issues in power systems, such as voltage stability, weak buses identification, abnormal and fault diagnosis. Combinations with model-based methods and data processing methods will improve the performance, and uncover more connections between electrical phenomena and statistical characteristics.

\appendices

\section{Random Matrix Theory}
\subsection{Ring Law}
Let $\widetilde{\mathbf X} \! \in \! \mathbb C^{N \times T}$ be a standard non-Hermitian random matrix, whose entries are independent identically distributed (i.i.d.) variables with
\begin{equation}
\mu(\widetilde{\mathbf x}_i)=0,\sigma^2(\widetilde{\mathbf x}_i)= 1 \quad (i=1,2,\ldots,N)
\end{equation}
where $\widetilde{\mathbf x}_i\!=\!(\widetilde x_{i,1},\widetilde x_{i,2},\!\ldots\!,\widetilde x_{i,T})$. For multiple standard non-Hermitian random matrices $\widetilde{\mathbf X}_i~(i\!=\!1,2,\!\dots\!,L)$, we define the matrix product as
\begin{equation}
\hat{\mathbf Z}=\prod_{i=1}^{L}\widetilde{\mathbf X}_{u,i}
\end{equation}
where $\widetilde{\mathbf X}_{u,i}$ is the singular value equivalent of $\widetilde{\mathbf X}_i$. The matrix product $\hat{\mathbf Z}$ can be converted to the standard matrix product $\widetilde{\mathbf Z}$, whose $\sigma^2(\widetilde{\mathbf z}_i)\!=\!\frac{1}{N}$ in each row. According to Ring Law \cite{guionnet2009single,benaych2013outliers}, the ESD of $\widetilde{\mathbf Z}$ converges almost surely to the limit with a probability density function (PDF) given by
\begin{eqnarray}
f_{\text{RL}}(\lambda_{\widetilde{\mathbf Z}})=\left\{
\begin{array}{ll}
\frac{1}{\pi c L}\vert \lambda_{\widetilde{\mathbf Z}} \vert ^{\frac{2}{L}-2} &(1-c)^{\frac{L}{2}} \leq \vert \lambda_{\widetilde{\mathbf Z}} \vert \leq 1\\
0 &\text{otherwise}
\end{array}
\right.
\end{eqnarray}
as $N,T\!\rightarrow\!\infty$ with a constant ratio $c\!=\!\frac{N}{T}\!\in\!(0,1]$. On the complex plane of eigenvalues, the inner circle radius is $(1\!-\!c)^\frac{L}{2}$ and the outer circle radius is unity.

\subsection{Marchenko-Pastur Law}
Let $\mathbf X\!=\!\{x_{i,j}\}_{N \times T}$ be a random matrix, whose entries are i.i.d. variables with
\begin{equation}
\mu(\mathbf x_i)=0,\sigma^2(\mathbf x_i)=d \quad (i=1,2,\ldots,N)
\end{equation}
where $d<\infty$ is a constant, and $\mathbf x_i\!=\!(x_{i,1},x_{i,2},\!\ldots\!,x_{i,T})$. The sample covariance matrix of $\mathbf X$ is defined as
\begin{equation}
\mathbf S=\frac{1}{N}\mathbf X\mathbf X^H
\end{equation}

According to M-P Law \cite{qiu2012cognitive,marvcenko1967distribution}, the ESD of $\mathbf S$ converges to the following PDF
\begin{eqnarray}
f_{\text{MP}}(\lambda_{\mathbf S})=\left\{
\begin{array}{ll}
\frac{1}{2\pi cd \lambda_{\mathbf S}} \sqrt{(b-\lambda_{\mathbf S})(\lambda_{\mathbf S}-a)} & a \leq \lambda_{\mathbf S} \leq b \\
0 &\text{otherwise}
\end{array}
\right.
\end{eqnarray}
as $N,T\!\rightarrow\!\infty$ with a constant ratio $c\!=\!\frac{N}{T}\!\in\!(0,1]$, where $a=d(1-\sqrt c)^2$, $b=d(1+\sqrt c)^2$.

\bibliographystyle{IEEEtran}
\bibliography{ShineRef}

\end{document}